\documentclass[10pt,final,twocolumn]{IEEEtran}

\ifCLASSINFOpdf
\else
\fi

\usepackage{graphicx}
\usepackage{subfigure}
\usepackage{amsmath}
\usepackage{dsfont}
\usepackage{algorithmic}
\usepackage{xspace}
\usepackage[ruled,vlined]{algorithm2e}
\usepackage{color}
\usepackage[font=small]{caption}
\usepackage{mathrsfs}
\usepackage{hhline}
\usepackage{enumitem}
\usepackage{multirow}

\usepackage{times}
\usepackage{psfrag}
\usepackage{amssymb,amsmath}
\usepackage[innercaption]{sidecap}
\usepackage{graphicx}
\usepackage{bm}
\usepackage{url}

\def\BI{\mbox{\tiny{BI}}}
\def\SO{{\mathrm{SO}}}
\def\SD{\mbox{\tiny{SD}}}
\def\BO{{\mathrm{BO}}}
\def\aBaseSuperframeDuration{{\emph{aBaseSuperframeDuration}}}
\def\aBaseSlotDuration{{\emph{aBaseSlotDuration}}}
\def\BE{{\mathrm{BE}}}
\def\CW{{\mathrm{CW}}}
\def\NB{{\mathrm{NB}}}

\def\Blue#1{\textcolor{black}{#1}}

\usepackage{array}
\newcolumntype{L}[1]{>{\raggedright\let\newline\\\arraybackslash\hspace{0pt}}m{#1}}
\newcolumntype{C}[1]{>{\centering\let\newline\\\arraybackslash}m{#1}}
\newcolumntype{R}[1]{>{\raggedleft\let\newline\\\arraybackslash\hspace{0pt}}m{#1}}
\newcolumntype{P}[1]{>{\centering\arraybackslash}p{#1}}

\begin{document}



\title{Wireless Network Design for Control Systems: \\ A Survey}


\author{Pangun Park, Sinem Coleri Ergen, Carlo Fischione, Chenyang Lu, and Karl Henrik Johansson \thanks{P. Park is with the Department of Radio and Information Communications Engineering, Chungnam National University, Korea (e-mail:~\texttt{pgpark@cnu.ac.kr}). S. Coleri Ergen is with the Department of Electrical and Electronics Engineering, Koc University, Istanbul, Turkey (e-mail:~\texttt{sergen@ku.edu.tr}). C. Lu is with the Department of Computer Science and Engineering, Washington University in St. Louis, St. Louis, USA (e-mail:~\texttt{lu@cse.wustl.edu}). C. Fischione and K. H. Johansson are with the ACCESS Linnaeus Center, Electrical Engineering, Royal Institute of Technology, Stockholm, Sweden (e-mail: \texttt{carlofi, kallej@ee.kth.se}). \textit{P. Park and S. Coleri Ergen contributed equally to this work.}}}


\IEEEcompsoctitleabstractindextext{
\begin{abstract}
Wireless networked control systems (WNCS) are composed of spatially distributed sensors, actuators, and controllers communicating through wireless networks instead of conventional point-to-point wired connections. Due to their main benefits in the reduction of deployment and maintenance costs, large flexibility and possible enhancement of safety, WNCS are becoming a fundamental infrastructure technology for critical control systems in automotive electrical systems, avionics control systems, building management systems, and industrial automation systems. The main challenge in WNCS is to jointly design the communication and control systems considering their tight interaction to improve the control performance and the network lifetime. In this survey, we make an exhaustive review of the literature on  wireless network design and optimization for WNCS. First, we discuss what we call the critical interactive variables including sampling period, message delay, message dropout, and network energy consumption. The mutual effects of these communication and control variables motivate their joint tuning. We discuss the effect of controllable wireless network parameters at all layers of the communication protocols on the probability distribution of these interactive variables. We also review the current wireless network standardization for WNCS and their corresponding methodology for adapting the network parameters. Moreover, we discuss the analysis and design of control systems taking into account the effect of the interactive variables on the control system performance. Finally, we present the state-of-the-art wireless network design and optimization for WNCS, while highlighting the tradeoff between the achievable performance and complexity of various approaches. We conclude the survey by highlighting major research issues and identifying future research directions.
\end{abstract}

}




\maketitle

\IEEEdisplaynotcompsoctitleabstractindextext
\IEEEpeerreviewmaketitle

\section{Introduction}
Recent advances in wireless networking, sensing, computing, and control are revolutionizing how control systems interact with information and physical processes such as Cyber-Physical Systems (CPS), Internet of Things (IoT), and \Blue{Tactile Internet}~\cite{Sztipanovits12, Bello16, Fettweis14}. In Wireless Networked Control Systems (WNCS), sensor nodes attached to the physical plant sample and transmit their measurements to the controller over a wireless channel; controllers compute control commands based on these sensor data, which are then forwarded to the actuators in order to influence the dynamics of the physical plant~\cite{Sadi14, Chen16}. \Blue{In particular, WNCS are strongly related to CPS and Tactile Internet since these emerging techniques deal with the real-time control of physical systems over the networks.} There is a strong technology push behind WNCS through the rise of embedded computing, wireless networks, advanced control, and cloud computing as well as a pull from emerging applications in automotive~\cite{Sadi13, ivwsn_4}, avionics~\cite{waic13}, building management~\cite{Witrant10_UFAD}, and industrial automation~\cite{Willig08,gungor09}. For example, WNCS play a key role in Industry 4.0~\cite{Ind40}. The ease of installation and maintenance, large flexibility, and increased safety make WNCS a fundamental infrastructure technology for safety-critical control systems. WNCS applications have been backed up by several international organizations such as Wireless Avionics Intra-Communications Alliance~\cite{waic13}, Zigbee Alliance~\cite{onworld_zigbee}, Z-wave Alliance~\cite{zwave}, International Society of Automation~\cite{isa_sp100}, Highway Addressable Remote Transducer communication foundation~\cite{whart_over}, and Wireless Industrial Networking Alliance~\cite{wisa}.

WNCS require novel design mechanisms to address the interaction between control and wireless systems for maximum overall system performance and efficiency. Conventional control system design is based on the assumption of instantaneous delivery of sensor data and control commands with extremely high reliabilities. The usage of wireless networks in the data transmission introduces non-zero delay and message error probability at all times. Transmission failures or deadline misses may result in the degradation of the control system performance, and even more serious economic losses or reduced human safety. Hence, control system design needs to include mechanisms to tolerate message loss and delay. On the other hand, wireless network design needs to consider the strict delay and reliability constraints of control systems. The data transmissions should be sufficiently reliable and deterministic with the latency on the order of seconds, or even milliseconds, depending on the time constraints of the closed-loop system~\cite{Willig08,gungor09}. Furthermore, removing cables for the data communication of sensors and actuators motivates the removal of the power supply to these nodes to achieve full flexibility. The limited stored battery or harvested energy of these components brings additional limitation on the energy consumption of the wireless network~\cite{Breath,Chang04, Ploennigs10}.

\begin{figure}[]\centering
  \subfigure[Control cost for various sampling periods and message loss probabilities.]
  {
    \psfrag{a}[][]{\footnotesize{\textbf{Maximum allowable control cost}}}
    \psfrag{b}[][]{\footnotesize{\textbf{Network constraints}}}
    \psfrag{X}[][]{\footnotesize{Sampling period (ms)}}
    \psfrag{Y}[][]{\footnotesize{Message loss probability}}
    \includegraphics[width = 0.8\columnwidth]{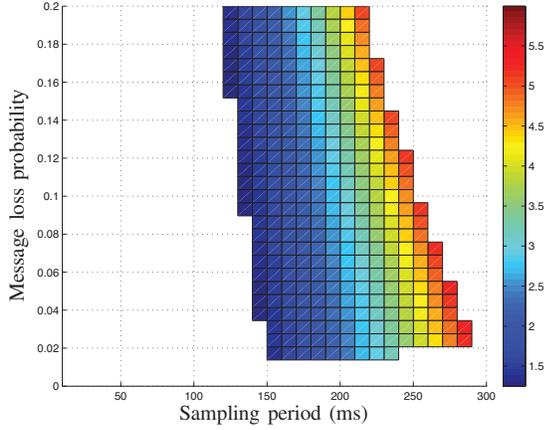}
    \label{fig:control-com1}
  }
  \subfigure[Control cost for various message delays and message loss probabilities.]
  {
  \psfrag{a}[][]{\footnotesize{\textbf{Maximum allowable control cost}}}
  \psfrag{b}[][]{\footnotesize{\textbf{Network constraints}}}
  \psfrag{X}[][]{\footnotesize{Message loss probability}}  
  \psfrag{Y}[][]{\footnotesize{Message delay (ms)}}
  \includegraphics[width = 0.8\columnwidth]{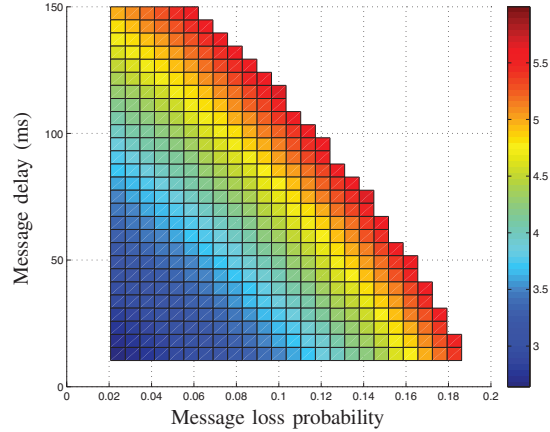}
  \label{fig:control-com2}
  }
 \caption{\Blue{Control cost of a WNCS using IEEE 802.15.4 protocol for various sampling periods, message delays and message loss probabilities.}}
 \label{fig:control-com}
 \end{figure}

The interaction between wireless networks and control systems can be illustrated by an example. A WNCS connects sensors attached to a plant to a controller via the single-hop wireless networking protocol IEEE 802.15.4. Fig.~\ref{fig:control-com} shows the control cost of the WNCS using the IEEE 802.15.4 protocol for different sampling periods, message delays and message loss probabilities~\cite{Pan11-thesis}. \Blue{The quadratic control cost is defined as a sum of the deviations of the plant state from its desired setpoint and the magnitude of the control input. The maximum allowable control cost is set to $6$.} The transparent region indicates that the maximum allowable control cost or network requirements are not feasible. For instance, the control cost would be minimized when there is no message loss and no delay, but this point is infeasible since these requirements cannot be met by the IEEE 802.15.4 protocol. The control cost generally increases as the message loss probability, message delay, and sampling period increase. Since short sampling periods increase the traffic load, the message loss probability, and the message delay are then closer to their critical values, above which the system is unstable~\cite{Schenato07}. Hence, the area and shape of the feasible region significantly depends on the network performance. Determining the optimal parameters for minimum network cost while achieving feasibility is not trivial because of the complex interdependence of the control and communication systems.






\Blue{Recently, Lower-Power Wide-Area Network (LPWAN) such as Long-Range WAN (LoRa)~\cite{LoRa} and NarrowBand IoT (NB-IoT)~\cite{NB_IoT} are developed to enable IoT connections over long-ranges (10--15~km). Even though some related works of WNCS are applicable for LPWAN-based control applications such as Smart Grid~\cite{Yu16}, Smart Transportation~\cite{Zhang11}, and Remote Healthcare~\cite{Marescaux01}, this survey focuses on wireless control systems based on Low-Power Wireless Personal Area Networks (LoWPAN) with short-range radios and their applications.} Some recent excellent surveys exist on wireless networks, particularly for industrial automation~\cite{Kumar14, Wang16, Lu16}. Specifically,~\cite{Kumar14} discusses the general requirements and representative protocols of Wireless Sensor Networks (WSNs) for industrial applications. \cite{Wang16} compares popular industrial WSN standards in terms of architecture and design. \cite{Lu16} mainly elaborates on real-time scheduling algorithms and protocols for WirelessHART networks, experimentation and joint wireless-control design approaches for industrial automation. While~\cite{Lu16} focused on WirelessHART networks and their control applications, this article provides a comprehensive survey of the design space of wireless networks for control systems and the potential synergy and interaction between control and communication designs. Specifically, our survey touches on the importance of interactions between recent advanced works of NCS and WSN, as well as different approaches of wireless network design and optimization for various WNCS applications.


\begin{figure}[]
  \centering
  \includegraphics[width = 1\columnwidth]{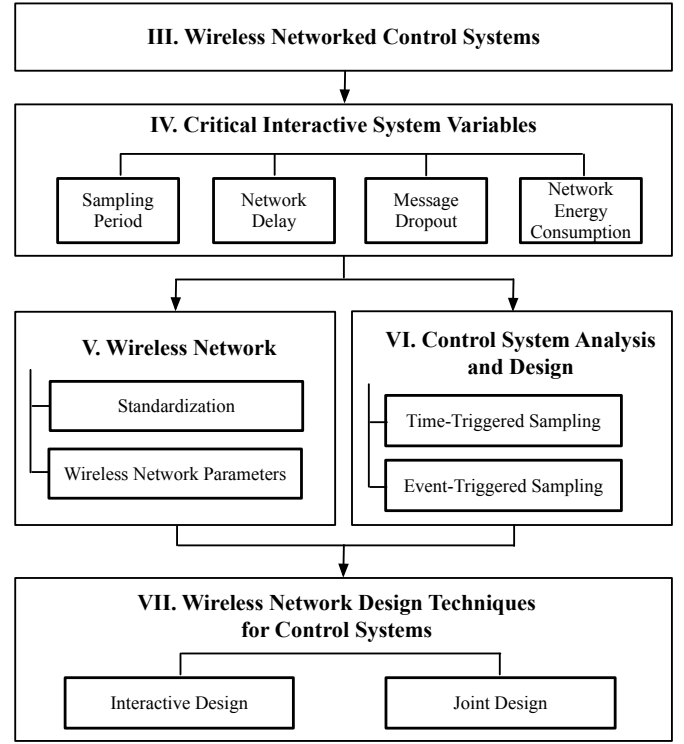} 
  \caption{Main section structure and relations. } \label{fig:diagram1}
\end{figure}

The goal of this survey is to unveil and address the requirements and challenges associated with wireless network design for WNCS and present a review of recent advances in novel design approaches, optimizations, algorithms, and protocols for effectively developing WNCS. The section structure and relations are illustrated in Fig.~\ref{fig:diagram1}. Section~\ref{sec:apps} introduces some inspiring applications of WNCS in automotive electronics, avionics, building automation, and industrial automation. Section~\ref{sec:wncs} describes WNCS where multiple plants are remotely controlled over a wireless network. Section~\ref{sec:para} presents the critical interactive variables of communication and control systems, including sampling period, message delay, message dropout, and energy consumption. Section~\ref{sec:net} introduces basic wireless network standardization and key network parameters at various protocol layers useful to tune the distribution of the critical interactive variables. Section~\ref{sec:control} then provides an overview of recent control design methods incorporating the interactive variables. Section~\ref{sec:design} presents various optimization techniques for wireless networks integrating the control systems. We classify the design approaches into two categories based on the degree of the integration: interactive designs and joint designs. In the interactive design, the wireless network parameters are tuned to satisfy given requirements of the control system. In the joint design, the wireless network and control system parameters are jointly optimized considering the tradeoff between their performances. \Blue{Section~\ref{sec:test} describes three experimental testbeds of WNCS.} We conclude this article by highlighting promising research directions in Section~\ref{sec:future}.

\section{Motivating Applications} \label{sec:apps}
This section explores some inspiring applications of WNCS.  

\subsection{Intra-Vehicle Wireless Network}
In-vehicle wireless networks have been recently proposed with the goal of reducing manufacturing and maintenance cost of a large amount of wiring harnesses within vehicles~\cite{Sadi13, ivwsn_4}. The wiring harnesses used for the transmission of data and power delivery within the current vehicle architecture may have up to $4\,000$ parts, weigh as much as $40$ kg and contain up to $4$ km of wiring. Eliminating these wires would additionally have the potential to improve fuel efficiency, greenhouse gas emission, and spur innovation by providing an open architecture to accommodate new systems and applications. 

An intra-vehicular wireless network consists of a central control unit, a battery, electronic control units, wireless sensors, and wireless actuators. Wireless sensor nodes send their data to the corresponding electronic control unit while scavenging energy from either one of the electronic control units or energy scavenging devices attached directly to them. Actuators receive their commands from the corresponding electronic control unit, and power from electronic control units or an energy scavenging device. The reason for incorporating energy scavenging into the envisioned architecture is to eliminate the lifetime limitation of fixed storage batteries. 

The applications that can exploit a wireless architecture fall into one of three categories: powertrain, chassis, and body. Powertrain applications use automotive sensors in engine, transmission, and onboard diagnostics for control of vehicle energy use, driveability, and performance. Chassis applications control vehicle handling and safety in steering, suspension, braking, and stability elements of the vehicle. Body applications include sensors mainly used for vehicle occupant needs such as occupant safety, security, comfort, convenience, and information. The first intra-vehicle wireless network applications are the Tire Pressure Monitoring System (TPMS)~\cite{tpms} and Intelligent Tire~\cite{intelligent_tire}. TPMS is based on the wireless transmission of tire pressure data from the in-tire sensors to the vehicle body. It is currently being integrated into all new cars in both U.S.A and Europe. Intelligent Tire is based on the placement of wireless sensors inside the tire to transfer accelerometer data to the coordination nodes in the body of the car with the goal of improving the performance of active safety systems. Since accelerometer data are generated at much higher rate than the pressure data and batteries cannot be placed within the tire, Intelligent Tire contains an ultra-low power wireless communication system powered by energy scavenging technology, which is now being commercialized by Pirelli~\cite{Pirelli}.

\subsection{Wireless Avionics Intra-Communication}
Wireless Avionics Intra-Communications (WAIC) have a tremendous potential to improve an aircraft's performance through more cost-effective flight operations, reduction in overall weight and maintenance costs, and enhancement of the safety~\cite{waic13}. Currently, the cable harness provides the connection between sensors and their corresponding control units to sample and process sensor information, and then among multiple control units over a backbone network for the safety-critical flight control~\cite{waic13,avi_req_92}. Due to the high demands on safety and efficiency, the modern aircraft relies on a large wired sensor and actuator networks that consist of more than $5\,000$ devices. Wiring harness usually represents 2--5\% of an aircraft's weight. For instance, the wiring harness of the Airbus A350-900 weights $23\,000$ kg~\cite{Oroitz10}.

The WAIC alliance considers wireless sensors of avionics located at various locations both within and outside the aircraft. The sensors are used to monitor the health of the aircraft structure, e.g., smoke sensors and ice detectors, and its critical systems, e.g., engine sensors and landing gear sensors. The sensor information is communicated to a central onboard entity. Potential WAIC applications are categorized into two broad classes according to application data rate requirements~\cite{waic_obj}. Low and high data rate applications have data rates less than and above 10~kbit/s, respectively. \

At the World Radio Conference 2015, the International Telecommunication Union voted to grant the frequency band 4.2--4.4~GHz for WAIC systems to allow the replacement of the heavy wiring used in aircraft~\cite{wrc15}. The WAIC alliance is dedicating efforts to the performance analysis of the assigned frequency band and the design of the wireless networks for avionics control systems~\cite{waic13}. Space shuttles and international space stations have already been using commercially available wireless solutions such as EWB MicroTAU and UltraWIS of Invocon~\cite{invocon}.



 

\subsection{Building Automation}
Wireless network based building automation provides significant savings in installation cost, allowing a large retrofit market to be addressed as well as new constructions. Building automation aims to achieve optimal level occupant comfort while minimizing energy usage~\cite{Aswani12}. These control systems are the integrative component to fans, pumps, heating/cooling equipment, dampers, and thermostats. The modern building control systems require a wide variety of sensing capabilities in order to control temperature, pressure, humidity, and flow rates. The European environment agency~\cite{Elec,water} shows that the electricity and water consumption of buildings are about $30\%$ and $43\%$ of the total resource consumptions, respectively. An On World survey~\cite{onworld_smart} reports that $59\%$ of 600 early adopters in five continents are interested in new technologies that will help them better manage their energy consumption, and $81\%$ are willing to pay for energy management equipment if they could save up to $30\%$ on their energy bill for smart energy home applications.

An example of energy management systems using WSNs is the intelligent building ventilation control described in~\cite{Witrant10_UFAD}. An underfloor air distribution indoor climate regulation process is set with the injection of a fresh airflow from the floor and an exhaust located at the ceiling level. The considered system is composed of ventilated rooms, fans, plenums, and wireless sensors. \Blue{A well-designed underfloor air distribution systems can reduce the energy consumption of buildings while improving the thermal comfort, ventilation efficiency and indoor air quality by using the low-cost WSNs.}

\subsection{Industrial Automation}
Wireless sensor and actuator network (WSAN) is an effective smart infrastructure for process control and factory automation~\cite{gungor09, Jiming10, Pister09}. Emerson Process Management~\cite{emerson} estimates that WSNs enable cost savings of up to 90\% compared to the deployment cost of wired field devices in the industrial automation domain. In industrial process control, the product is processed in a continuous manner (e.g., oil, gas, chemicals). In factory automation or discrete manufacturing, instead, the products are processed in discrete steps with the individual elements (e.g., cars, drugs, food). Industrial wireless sensors typically report the state of a fuse, heating, ventilation, or vibration levels on pumps. Since the discrete product of the factory automation requires sophisticated operations of robot and belt conveyors at high speed, the sampling rates and real-time requirements are often stricter than those of process automation. Furthermore, many industrial automation applications might in the future require battery-operated networks of hundreds of sensors and actuators communicating with access points.


According to TechNavio~\cite{technavio}, WSN solutions in industrial control applications is one of the major emerging industrial trends. Many wireless networking standards have been proposed for industrial processes, e.g., WirelessHART by ABB, Emerson, and Siemens and ISA~100.11a by Honeywell~\cite{Petersen11}. Some industrial wireless solutions are also commercially available and deployed such as Tropos of ABB and Smart Wireless of Emerson.

\begin{figure}[]
  \centering
  \includegraphics[width = 1.0\columnwidth]{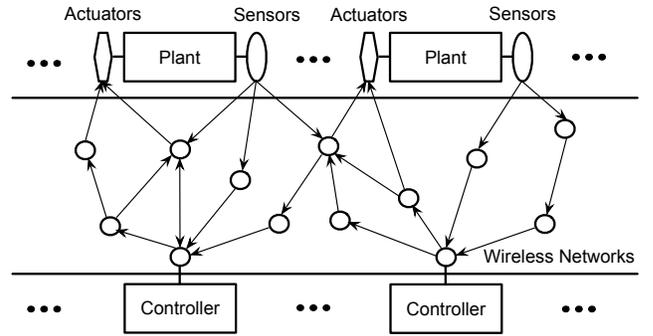}
  \caption{Overview of the considered NCS setup. Multiple plants are controlled by multiple controllers. A wireless network closes the loop from sensor to controller and from controller to actuator. The network includes not only nodes attached to the plant or controller, but also relay nodes. \label{fig:wsan}}
\end{figure}

\section{Wireless Networked Control Systems}\label{sec:wncs}
Fig.~\ref{fig:wsan} depicts the generalized closed-loop diagram of WNCS where multiple plants are remotely controlled over a wireless network~\cite{Joao07}. The wireless network includes sensors and actuators attached to the plants, controllers, and relay nodes. A plant is a continuous-time physical system to be controlled. The inputs and outputs of the plant are continuous-time signals. Outputs of plant $i$ are sampled at periodic or aperiodic intervals by the wireless sensors. Each packet associated to the state of the plant is transmitted to the controller over a wireless network. When the controller receives the measurements, it computes the control command. The control commands are then sent to the actuator attached to the plant. Hence, the closed-loop system contains both a continuous-time and a sampled-data component. Since both sensor--controller and controller--actuator channels use a wireless network, general WNCS of Fig.~\ref{fig:wsan} are also called two-channel feedback NCS~\cite{Joao07}. The system scenario is quite general, as it applies to any interconnection between a plant and a controller.

\subsection{\Blue{Control Systems}}
The objective of the feedback control system is to ensure that the closed-loop system has desirable dynamic and steady-state response characteristics, and that it is able to efficiently attenuate disturbances and handle network delays and loss. Generally, the closed-loop system should satisfy various design objectives: stability, fast and smooth responses to set-point changes, elimination of steady-state errors, avoidance of excessive control actions, and a satisfactory degree of robustness to process variations and model uncertainty~\cite{Astrom97}. \Blue{In particular, the stability of a control system is an extremely important requirement. Most NCS design methods consider subsets of these requirements to synthesize the estimator and the controller. In this subsection, we briefly introduce some fundamental aspects of modeling, stability, control cost, and controller and estimator design for NCSs.}

\subsubsection{NCS Modeling}
\Blue{NCSs can be modeled using three main approaches, namely, the discrete-time approach, the sampled-data approach, and the continuous-time approach, dependent on the controller and the plant~\cite{NCS10}. The discrete-time approach considers discrete-time controllers and a discrete-time plant model. The discrete-time representation leads often to an uncertain discrete-time system in which the uncertainties appear in the matrix exponential form due to discretization. Typically, this approach is applied to NCS with linear plants and controllers since in that case exact discrete-time models can be derived.}

\Blue{Secondly, the sampled-data approach considers discrete-time controllers but for a continuous-time model that describes the sampled-data NCS dynamics without exploiting any form of discretization~\cite{SDS95}. Delay-differential equations can be used to model the sampled-data dynamics. This approach is able to deal simultaneously with time-varying delays and time-varying sampling intervals.}

\Blue{Finally, the continuous-time approach designs a continuous-time controller to stabilize a continuous-time plant model. The continuous-time controller then needs to be approximated by a representation suitable for computer implementation~\cite{Astrom97}, whereas typical WNCS consider the discrete-time controller. We will discuss more details of the analysis and design of WNCS to deal with the network effects in Section~\ref{sec:control}.}

\subsubsection{Stability}
\Blue{Stability is a base requirement for controller design. We briefly describe two fundamental notions of stability, namely, input-output stability and internal stability~\cite{CS08}. While the input-output stability is the ability of the system to produce a bounded output for any bounded input, the internal stability is the system ability to return to equilibrium after a perturbation. For linear systems, these two notions are closely related, but for nonlinear system they are not the same.}

\Blue{Input-output stability concerns the forced response of the system for a bounded input. A system is defined to be \textit{Bounded-Input-Bounded-Output (BIBO) stable} if every bounded input to the system results in a bounded output. If for any bounded input the output is not bounded the system is said to be \textit{unstable}.}


\Blue{Internal stability is based on the magnitude of the system response in steady state. If the steady-state response is unbounded, the system is said to be \textit{unstable}. A system is said to be \textit{asymptotically stable} if its response to any initial conditions decays to zero asymptotically in the steady state. A system is defined to be \textit{exponentially stable} if the system response in addition decays exponentially towards zero. The faster convergence often means better performance. In fact, many NCS researches analyze exponential stability conditions~\cite{Zhang01, Walsh02}. Furthermore, if the response due to the initial conditions remains bounded but does not decay to zero, the system is said to be \textit{marginally stable}. Hence, a system cannot be both asymptotically stable and marginally stable. If a linear system is asymptotically stable, then it is BIBO stable. However, BIBO stability does not generally imply internal stability. Internal stability is stronger in some sense, because BIBO stability can hide unstable internal behaviors, which do not appear in the output.}



\subsubsection{Control Cost}
\Blue{Besides stability guarantees, typically a certain closed-loop control performance is desired. The closed-loop performance of a control system can be quantified by the control cost as a function of plant state and control inputs~\cite{CS08}. A general regulation control goal is to keep the state error from the setpoint close to zero, while minimizing the control actions. Hence, the control cost often consists of two terms, namely, the deviations of plant state from their desired setpoint and the magnitude of the control input. A common controller design approach is via a Linear Quadratic control formulation for linear systems and a quadratic cost function~\cite{Athans71}. The quadratic control cost is defined as a sum of the quadratic functions of the state deviation and the control effort. In such formulation, the optimal control policy that minimizes the cost function can be explicitly computed from a Riccati equation.}





\subsubsection{Controller Design}
\Blue{The controller should ensure that the closed-loop system has desirable dynamic and steady state response characteristics. For NCS, the network delay and loss may degrade the control performance and even destabilize the system. Some surveys present controller design for NCSs~\cite{Joao07, Tipsuwan03}. For a historical review, see the survey~\cite{Astrom14}. We briefly describe three representative controllers, namely, Proportional-Integral-Derivative (PID) controller~\cite{Astrom1995a}, Linear Quadratic Regulator (LQR) control~\cite{Athans71}, and Model Predictive Control (MPC)~\cite{Qin03}.}

\Blue{PID control is almost a century old and has remained the most widely used controller in process control until today~\cite{Astrom1995a}. One of the main reasons for this controller to be so widely used is that it can be designed without precise knowledge of the plant model. A PID controller calculates an error value as the difference between a desired setpoint and a measured plant state. The control signal is a sum of three terms: the P-term (which is proportional to the error), the I-term (which is proportional to the integral of the error), and the D-term (which is proportional to the derivative of the error). The controller parameters are proportional gain, integral time, and derivative time. The integral, proportional, and derivative part can be interpreted as control actions based on the past, the present and the future of the plant state. Several parameter tuning methods for PID controllers exist~\cite{Astrom1995a, Barbosa04}. Historically, PID tuning methods require a trial and error process in order to achieve a desired stability and control performance.}


\Blue{The linear quadratic problem is one of the most fundamental optimal control problems where the objective is to minimize a quadratic cost function subject to plant dynamics described by a set of linear differential equations~\cite{Athans71}. The quadratic cost is a sum of the plant state cost, final state cost, and control input cost. The optimal controller is a linear feedback controller. The LQR algorithm is basically an automated way to find the state-feedback controller. Furthermore, the LQR is an important subproblem of the general Linear Quadratic Gaussian (LQG) problem. The LQG problem deals with uncertain linear systems disturbed by additive Gaussian noise.  While the LQR problem assumes no noise and full state observation, the LQG problem considers input and measurement noise and partial state observation.}



\Blue{Finally, MPC solves an optimal linear quadratic control problems over a receding horizon~\cite{Qin03}. Hence, the optimization problem is similar to the controller design problem of LQR but solved over a moving horizon in order to handle model uncertainties. In contrast to non-predictive controllers, such as a PID or a LQR controller, which compute the current control action as a function of the current plant state using the information about the plant from the past, predictive controllers compute the control based on the systems predicted future behaviour~\cite{Carlos89}. MPC tries to optimize the system behaviour in a receding horizon fashion. It takes control commands and sensing measurements to estimate the current and future state of plant based on the control system model. The control command is optimized to get the desired plant state based on a quadratic cost. In practice, there are often hard constraints imposed on the state and the control input. Compared to the PID and LQR control, the MPC framework efficiently handles constraints. Moreover, MPC can handle missing measurements or control commands~\cite{Henriksson15, Li16}, which can appear in a NCS setting.}


\subsubsection{Estimator Design}
\Blue{Due to network uncertainties, plant state estimation is a crucial and significant research field of NCSs~\cite{Joao07, Schenato07}. An estimator is used to predict the plant state by using partially received plant measurements. Moreover, the estimator typically compensates measurement noise, network delays, and packet losses. This predicted state is sometimes used in the calculation of the control command. Kalman filter is one of the most popular approaches to obtain the estimated plant states for NCS~\cite{Sinopoli04}. Modified Kalman filters are proposed to deal with different models of the network delay and loss~\cite{Schenato07, Li16, Sahebsara07, burak_opt}. The state estimation problem is often formulated by probabilistically modeling the uncertainties occurring between the sensor and the controller~\cite{Schenato07, Sinopoli04, Sahebsara07, Schenato08}. However, a non-probabilistic approach by time-stamping the measurement packets is proposed in~\cite{Moayedi11}.}


\Blue{In LQG control, a Kalman filter is used to estimate the state from the plant output. The optimal state estimator and the optimal state feedback controller are combined for the LQG problem. The controller is the linear feedback controller of LQR. The optimal LQG estimator and controller can be designed separately if the communication protocol supports the acknowledgement of the packet transmission of both sensor--controller and controller--actuator channels~\cite{Schenato07}. In sharp contrast, the separation principle between estimator and controller does not hold if the acknowledgement is not supported~\cite{Maben16}. Hence, the underlying network operation is critical in the design of the overall estimator and the controller.}

%

\subsection{Wireless Networks}

For the vast majority of control applications, most of the traffic over the wireless network consists of real-time sensor data from sensor nodes towards one or more controllers. The controller either sits on the backbone or is reachable via one or more backbone access points. Therefore, data flows between sensor nodes and controllers are not necessarily symmetric in WNCS. In particular, asymmetrical link cost and unidirectional routes are common for the most part of the sensor traffic. Furthermore, multiple sensors attached to a single plant may independently transmit their measurements to the controller~\cite{Matveev03}. In some other process automation environments, multicast may be used to deliver data to multiple nodes that may be functionally similar, such as the delivery of alerts to multiple nodes in an automation control room. 

Wireless sensors and actuators in control environments can be powered by battery, energy scavenging, or power cable. Battery storage provides a fixed amount of energy and requires replacement once the energy is consumed. Therefore, efficient usage of energy is vital in achieving high network lifetime. Energy harvesting techniques, on the other hand, may rely on natural sources, such as solar, indoor lighting, vibrational, thermal~\cite{natural_harvesting}, inductive and magnetic resonant coupling~\cite{coupling}, and radio frequency~\cite{RF_harvesting}. Efficient usage of energy harvesting may attain infinite lifetime for the sensor and actuator nodes. In most situations, the actuations need to be powered separately because significant amount of energy is required for the actuation commands (e.g., opening a valve).

\begin{figure}[]\centering
  \psfrag{x}[][]{\footnotesize{\textbf{Maximum allowable control cost}}}
  \psfrag{y}[][]{\footnotesize{\textbf{Network constraints}}}
  \psfrag{a}[][]{\footnotesize{Message delay (ms)}}
  \psfrag{b}[][]{\footnotesize{Packet loss probability}}
  \psfrag{c}[][]{\footnotesize{Sampling period (ms)}}
  \includegraphics[width = 1\columnwidth]{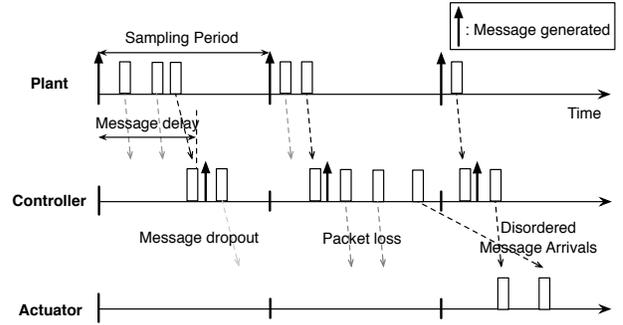}
  \caption{Timing diagram for closed-loop control over a wireless network with sampling period, message delay, and message dropouts.}
  \label{fig:time}
\end{figure}

\section{Critical Interactive System Variables}\label{sec:para}
The critical system variables creating interactions between WNCS control and communication systems are sampling period, message delay, and message dropout. Fig.~\ref{fig:time} illustrates the timing diagram of the closed-loop control over a wireless network with sampling period, message delay, and message dropouts. \Blue{We distinguish \textit{messages} of the control application layer with \textit{packets} of the communication layer. The control system generates messages such as the sensor samples of the sensor--controller channel or the control commands of the controller--actuator channel. The control system generally determines the sampling period. The communication protocols then convert the message to the packet format and transmit the packet to the destination. Since the wireless channel is lossy, the transmitter may have multiple packet retransmissions associated to one message depending on the communication protocol.} If all the packet transmissions of the message fail due to a bursty channel, then the message is considered to be lost. 



In Fig.~\ref{fig:time}, the message delay is the time delay between when the message was generated by the control system at a sensor or a controller and when it is received at the destination. Hence, the message delay of a successfully received message depends on the number of packet retransmissions. Furthermore, since the routing path or network congestion affects the message delay, the message arrivals are possibly disordered as shown in Fig.~\ref{fig:time}.

\begin{figure*}[t]\centering
  \psfrag{x}[][]{\footnotesize{\textbf{Maximum allowable control cost}}}
  \psfrag{y}[][]{\footnotesize{\textbf{Network constraints}}}
  \psfrag{a}[][]{\footnotesize{Message delay (ms)}}
  \psfrag{b}[][]{\footnotesize{Packet loss probability}}
  \psfrag{c}[][]{\footnotesize{Sampling period (ms)}}
  \includegraphics[width = 1.6\columnwidth]{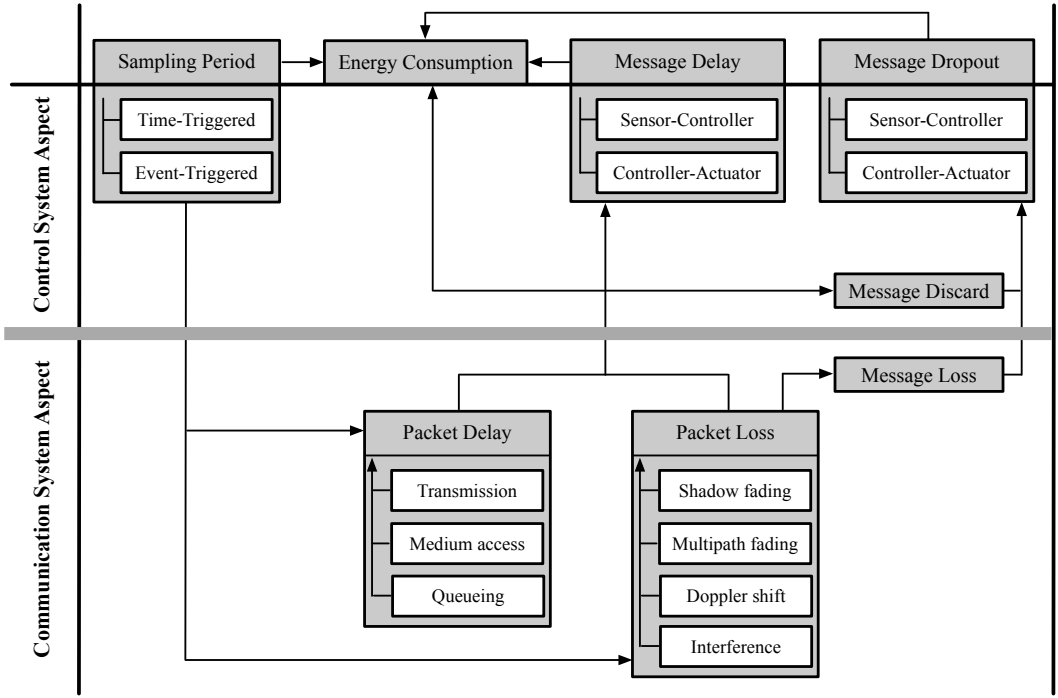}
  \caption{\Blue{Complex interactions between critical system variables. The arrows represent some of the explicit relationships.}}
  \label{fig:inter}
\end{figure*}

The design of the wireless network at multiple protocol layers determines the probability distribution of message delay and message dropout. These variables together with the sampling period influence the stability of the closed-loop NCS and the energy consumption of the network. Fig.~\ref{fig:inter} presents the dependences between the critical system variables. Since WNCS design requires an understanding of the interplay between communication and control, we discuss the effect of these system variables on both control and communication system performance.

\subsection{Sampling Period}

\subsubsection{Control System Aspect}
Continuous-time signals of the plant need to be sampled before they are transmitted through a wireless network. It is important to note that the choice of the sampling should be related to the desired properties of the closed-loop system such as the response to reference signals, influence of disturbances, network traffic, and computational load~\cite{wit02}. There are two methods to sample continuous-time signals in WNCS: \textit{time-triggered} and \textit{event-triggered sampling}~\cite{self-event}.

In time-triggered sampling, the next sampling instant occurs after the elapse of a fixed time interval, regardless of the plant state. Periodic sampling is widely used in digital control systems due to the simple analysis and design of such systems. Based on experience and simulations, a common rule for the selection of the sampling period is to make sure $\omega \, h $ be in the range $[0.1,0.6]$ , where $\omega$ is the desired natural frequency of the closed-loop system and $h$ is the sampling period~\cite{wit02}. This implies typically that we are sampling up to $20$ samples per period of the dominating mode of the closed-loop system.

In a traditional digital control system based on point-to-point wired connections, the smaller the sampling period is chosen, the better the performance is achieved for the control system~\cite{Franklin97}. However, in wireless networks, the decrease in sampling period increases the network traffic, which in turn increases the message loss probability and message delay. Therefore, the decrease in sampling period eventually degrades the control performance, as illustrated in Fig.~\ref{fig:control-com}.


Recently, event-based control schemes such event- and self-triggered control systems have been proposed, where sensing and actuating are performed when the system needs attention~\cite{self-event}. Hence, the traffic pattern of event- and self-triggered control systems is asynchronous rather than periodic. In event-triggered control, the execution of control tasks is determined by the occurrence of an event rather than the elapse of a fixed time period as in time-triggered control. Events are triggered only when stability or a pre-specified control performance are about to be lost~\cite{Tabuada07,Lemmon09,Lunze10}. Event-triggered control can significantly reduce the traffic load of the network with no or minor control performance degradation since the traffic is generated only if the signal changes by a specified amount~\cite{Arzen99, Jose14}. However, since most trigger conditions depend on the instantaneous state, the plant state is required to be monitored~\cite{Tabuada07, Lunze10}. Self-triggered control has been proposed to prevent such monitoring~\cite{Wang09}. In self-triggered control, an estimation of the next event time instant is made. The online detection of plant disturbances and corresponding control actions cannot be generated with self-triggered control. A combination of event- and self-triggered control is therefore often desirable~\cite{Jose14, Peng16}.



 \subsubsection{Communication System Aspect}
The choice of time-triggered and event-triggered sampling in the control system determines the pattern of message generation in the wireless network. Time-triggered sampling results in regular periodic message generation at predetermined rate. If random medium access mechanism is used, the increase in network load results in worse performance in the other critical interactive system variables, i.e., message delay, message dropout, and energy consumption~\cite{sinem_analysis}. The increase in control system performance with higher sampling rates, therefore, does not hold due to these network effects. On the other hand, the predetermined nature of packet transmissions in time-triggered sampling allows explicit scheduling of sensor node transmissions beforehand, reducing the message loss and delay caused by random medium access~\cite{pedamacs, sinem_tdma}.  A scheduled access mechanism can predetermine the transmission time of all the components such that additional nodes have minimal effect on the transmission of existing nodes~\cite{Sadi13, sinem_twc}. When the transmission of the periodically transmitting nodes are distributed uniformly over time rather than being allocated immediately as they arrive, additional nodes may be allocated without causing any jitter in their periodic allocation.

The optimal choice of medium access control mechanism is not trivial for event-triggered control~\cite{Jose14, Blind11_1}. The overall performance of event-triggered control systems significantly depends on the plant dynamics and the number of control loops. The random access mechanism is a good alternative if a large number of slow dynamical plants share the wireless network. In this case, the scheduled access mechanism may result in significant delay between the triggering of an event and a transmission in its assigned slot due to the large number of control loops. However, most time slots are not utilized since the traffic load is low for slow plants. On the other hand, the scheduled access mechanism performs well when a small number of the fast plants is controlled by the event-triggered control algorithm. Contention-based random access generally degrades the reliability and delay performance for the high traffic load of fast plants. When there are packet losses in the random access scheme, the event-triggered control further increases the traffic load, which may eventually incur stability problems~\cite{Blind11_1}.

The possible event-time prediction of self-triggered control alleviates the high network load problem of time-triggered sampling and random message generation nature of event-triggered sampling by predicting the evolution of the triggering threshold crossings of the plant state~\cite{self-event}. The prediction allows the explicit scheduling of sensor node transmissions, eliminating the high message delays and losses of random medium access. Most existing works of event-triggered and self-triggered control assume that message dropouts and message disorders do not occur. This assumption is not practical when the packets of messages are transmitted through a wireless network. Dealing with message dropouts and message disorders in these control schemes is challenging for both the wireless network and the control system.



\subsection{Message Delay}

\subsubsection{Control System Aspect}
There are mainly two kinds of message delays of NCSs: sensor--controller delay and controller--actuator delay, as illustrated in Fig.~\ref{fig:time}. The sensor--controller delay represents the time interval from the instant when the physical plant is sampled to the instant when the controller receives the sampled message; and the controller--actuator delay indicates the time duration from the generation of the control message at the controller until its reception at the actuator. The increase in both delays prevents the timely delivery of the control feedback, which degrades system performance, as exemplified in Fig.~\ref{fig:control-com}. In control theory, these delays cause phase shifts that limit the control bandwidth and affect closed-loop stability~\cite{wit02}. 


Since delays are especially pernicious for closed-loop systems, some forms of modeling and prediction are essential to overcome their effects. Techniques proposed to overcome sensor--controller delays use predictive filters including Kalman filter~\cite{wit02, smith57, Sinopoli04, Sahebsara07}. In practice, message delay can be estimated from time stamped data if the receiving node is synchronized through the wireless network~\cite{isa_sp100, whart_over}. The control algorithm compensates the measured or predicted delay unless it is too large~\cite{smith57}. Such compensation is generally impossible for controller--actuator delays. Hence, controller--actuator delays are more critical than the sensor--controller delays~\cite{Schenato07,Joao07}.


The packet delay variation is another interesting metric since it significantly affects the control performance and causes possible instability even when the mean delay is small. In particular, a heavy tail of the delay distribution significantly degrades the stability of the closed-loop system~\cite{Seuret12}. The amount of degradation depends on the dynamics of the process and the distribution of the delay variations. One way to eliminate delay variations is to use a buffer, trading delay for its variation.

\subsubsection{Communication System Aspect}
Message delay in a multihop wireless network consists of transmission delay, access delay, and queueing delay at each hop in the path from the source to the destination. 
 
Transmission delay is defined as the time required for the transmission of the packet. Transmission delay depends on the amount of data to be transmitted to the destination and the transmission rate, which depends on the transmit power of the node itself and its simultaneously active neighboring nodes. As the transmit power of the node increases, its own transmission rate increases, decreasing its own transmission delay; while causing more interference to simultaneously transmitting nodes, increasing their delay. The optimization of transmission power and rate should take into account this tradeoff~\cite{mindelay}.

Medium access delay is defined as the time duration required to start the actual transmission of the packet. Access delay depends on the choice of medium access control (MAC) protocol. If contention-based random access mechanism is used, this delay depends on the network load, encoding/decoding mechanism used in the transmitter and receiver, and random access control protocol. As the network load increases, the access delay increases due to the increase in either busy sensed channel or failed transmissions. The receiver decoding capability determines the number of simultaneously active neighboring transmitters. The decoding technique may be based on interference avoidance, in which only one packet can be received at a time~\cite{mindelay}; self-interference cancellation, where the node can transmit another packet while receiving~\cite{interf-self}; or interference cancellation, where the node may receive multiple packets simultaneously and eliminate interference~\cite{interf_sic}. Similarly, a transmitter may have the capability to transmit multiple packets simultaneously~\cite{uysal}. The execution of the random access algorithm together with its parameters also affect the message delay. On the other hand, if schedule-based access is used, the access delay in general increases as the network load increases. However, this effect may be minimized by designing efficient scheduling algorithms adopting uniform distribution of transmissions via exploiting the periodic transmission of time-triggered control~\cite{Sadi13, sinem_twc}. Similar to random access, more advanced encoding/decoding capability of the nodes may further decrease this access delay. Moreover, packet losses over the channel may require retransmissions, necessitating the repetition of medium access and transmission delay over time. This further increases message delay, as illustrated in Fig.~\ref{fig:time}.

Queueing delay depends on the message generation rate at the nodes and amount of data they are relaying in the multihop routing path. The message generation and forwarding rate at the nodes should be kept at an acceptable level so as not to allow packet build up at the queue. Moreover, scheduling algorithms should consider the multihop forwarding in order to minimize the end-to-end delay from the source to the destination~\cite{burak_opt, sinem_tdma, baldi09}. The destination may observe disordered messages since the packet associated to the message travels several hops with multiple routing paths or experiences network congestion~\cite{isa_sp100, whart}.



\subsection{Message Dropout}


\begin{table*}
\center
\scriptsize
\begin{tabular}{|C{1.9cm} || C{2.7cm} | C{3.1cm} | C{3.5cm} | C{4.0cm} | } 
 \hline
{} & Physical Layer & Medium Access Control & Data Link Layer & Routing \\ 
\hhline{|=#=|=|=|=|} 
IEEE 802.15.4 & DSSS & CSMA/CA, GTS allocation & - & -\\ 
 \hline
WirelessHART & IEEE 802.15.4 PHY & IEEE 802.15.4 MAC & TDMA, Channel hoping, Channel blacklisting & Source routing, Graph routing  \\ 
 \hline
ISA-100.11a & IEEE 802.15.4 PHY & IEEE 802.15.4 MAC & TDMA, Channel hoping, Channel blacklisting & Source routing, Graph routing  \\ 
 \hline
IEEE 802.15.4e & IEEE 802.15.4 PHY & TSCH, DSME, LLDN & - & - \\ 
 \hline
6LoWPAN & IEEE 802.15.4 PHY & IEEE 802.15.4 MAC & Compaction, Fragmentation & - \\ 
 \hline
RPL & Any & Any & - & Source routing, Distance vector routing \\ 
 \hline
6TiSCH & IEEE 802.15.4 PHY & TSCH & Management, Resource allocation, Performance monitoring & - \\ 
 \hline
IEEE 802.11 & DSSS, OFDM & DCF, PCF & - & - \\ 
 \hline
IEEE 802.11e & DSSS, OFDM & EDCA, HCCA & - & - \\ 
 \hline 
\end{tabular}
\caption{\Blue{Comparison of wireless standards}} \label{tab:standard_comparison}
\end{table*}

\subsubsection{Control System Aspect}
Generally, there are two main reasons for message dropouts, namely, message discard due to the control algorithm and message loss due to the wireless network itself. The logical Zero-Order Hold (ZOH) mechanism is one of the most popular and simplest approaches to discard disordered messages~\cite{Joao07, Xiong09, Cloosterman10}. In this mechanism, the latest message is kept and old messages are discarded based on the time stamp of the messages. However, some alternatives are also proposed to utilize the disordered messages in a filter bank~\cite{Moyne07, Nilsson98-thesis}. A message is considered to be lost if all packet transmissions associated to the message have eventually failed. The effect of message dropouts is more critical than message delay since it increases the updating interval with a multiple of the sampling period.

There are mainly two types of dropouts: sensor--controller message dropouts and controller--actuator message dropouts. The controller estimates the plant state to compensate possible message dropouts of the sensor--controller channel. Remind that Kalman filtering is one of the most popular approaches to estimate the plant state and works well if there is no significant message loss~\cite{Schenato07}. Since the control command directly affects the plant, controller--actuator dropouts are more critical than sensor--controller dropouts~\cite{Park15_tr, Saifullah14_rate}. Many practical NCSs have several sensor--controller channels whereas the controllers are collocated with the actuators, e.g., heat, ventilation and air-conditioning control systems~\cite{Arampatzis06}.

NCS literatures often model the message dropout as a stochastic variable based on different assumptions of the maximum consecutive message dropouts. In particular, significant work has been devoted for deriving upper bounds on the updating interval for which stability can be guaranteed~\cite{Branicky00, Zhang01, Zhang09}. The upper bounds could be used as the update deadline over the network as we will discuss in more detail in Section~\ref{sec:control}. The bursty message dropout is very critical for control systems since it directly affects the upper bounds on the updating interval. 






\subsubsection{Communication System Aspect}
Data packets may be lost during their transmissions, due to the susceptibility of wireless channel to blockage, multipath, doppler shift, and interference~\cite{Goldsmith05}. Obstructions between transmitter and receiver, and their variation over time, cause random variations in the received signal, called shadow fading. The probabilistic distribution of the shadow fading depends on the number, size, and material of the obstructions in the environment. Multipath fading, mainly caused by the multipath components of the transmitted signal reflected, diffracted or scattered by surrounding objects, occurs over shorter time periods or distances than shadow fading. The multipath components arriving at the receiver cause constructive and destructive interference, changing rapidly over distance. Doppler shift due to the relative motion between the transmitter and the receiver may cause the signal to decorrelate over time or impose lower bound on the channel error rate. Furthermore, unintentional interference from the simultaneous transmissions of neighboring nodes and intentional interference in the form of cyber-attacks can disturb the successful reception of packets as well.



\subsection{Network Energy Consumption} 
A truly wireless solution for WNCS requires removing power cables in addition to the data cables to provide full flexibility of installation and maintenance. Therefore, the nodes need to rely on either battery storage or energy harvesting techniques. Limiting the energy consumption in the wireless network prolongs the lifetime of the nodes. If enough energy scavenging can be extracted from natural sources, inductive or magnetic resonant coupling, or radio frequency, then infinite lifetime may be achieved~\cite{natural_harvesting,RF_harvesting}.


Decreasing sampling period, message delay, and message dropout improves the performance of the control system, but at the cost of higher energy consumption in the communication system~\cite{Park12-adaptive}. The higher the sampling rate, the greater the number of packets to be transmitted over the channel. This increases the energy consumption of the nodes. Moreover, decreasing message delay requires increasing the transmission rate or data encoding/decoding capability at the transceivers. This again comes at the cost of increased energy consumption~\cite{uysal_energy}. Finally, decreasing message dropout requires either increasing transmit power to combat fading and interference, or increasing data encoding/decoding capabilities. This again translates into energy consumption.
 

\section{Wireless Network}\label{sec:net}

\subsection{Standardization}\label{sec:standard}
\Blue{The most frequently adopted communication standards for WNCS are IEEE 802.15.4 and IEEE 802.11 with some enhancements. Particularly, WirelessHART, ISA-100.11a, and IEEE 802.15.4e are all based on IEEE 802.15.4. Furthermore, some recent works of IETF consider Internet Protocol version 6 (IPv6) over low-power and lossy networks such as 6LoWPAN, Routing Protocol for Low-Power and Lossy Networks (RPL), and 6TiSCH, which are all compatible with IEEE 802.15.4~\cite{Palattella13}.} IEEE 802.15.4 is originally developed for low-rate, low-power and low-cost Personal Area Networks (PANs) without any concern on delay and reliability. The standards such as WirelessHART, ISA-100.11a and IEEE 802.15.4e are built on top of the physical layer of IEEE 802.15.4 with additional Time Division Multiple Access (TDMA), frequency hopping and multiple path features to provide delay and reliable packet transmission guarantees while further lowering energy consumption. \Blue{In this subsection, we first introduce IEEE 802.15.4 and then discuss WirelessHART, ISA-100.11a, IEEE 802.15.4e, and the higher layers of IETF activities such as 6LoWPAN, RPL, and 6TiSCH.} 

\Blue{On the other hand, although the key intentions of the IEEE 802.11 family of Wireless Local Area Network (WLAN) standards are to provide high throughput and a continuous network connection, several extensions have been proposed to support QoS for wireless industrial communications~\cite{Willig02, Tian12}. In particular, the IEEE 802.11e specification amendment introduces significant enhancements to support the soft real-time applications. In this subsection, we will describe the fundamental operations of basic IEEE 802.11 and IEEE 802.11e.} \Blue{The standards are summarized in Table~\ref{tab:standard_comparison}.}


\subsubsection{IEEE 802.15.4}\label{sec:overview-wpan}
\Blue{IEEE 802.15.4 standard defines the physical and MAC layers of the protocol stack~\cite{ieee802154}. A PAN consists of a PAN coordinator that is responsible of managing the network and many associated nodes. The standard supports both star topology, in which all the associated nodes directly communicate with the PAN coordinator, and peer-to-peer topology, where the nodes can communicate with any neighbouring node while still being managed by the PAN coordinator.} 

\Blue{The physical layer adopts direct sequence spread spectrum, which is based on spreading the transmitted signal over a large bandwidth to enable greater resistance to interference. A single channel between $868$ and $868.6$ MHz, $10$ channels between $902.0$ and $928.0$ MHz, and 16 channels between $2.4$ and $2.4835$ GHz are used. The transmission data rate is $250$ kbps in the $2.4$ GHz band, $40$ kbps in $915$ MHz and $20$ kbps in $868$ MHz band.}

\begin{figure}[t]
    \centering
    \includegraphics[width = 1\columnwidth]{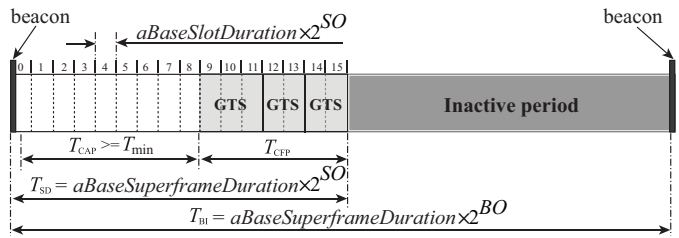}
    \caption{\Blue{Superframe structure of IEEE 802.15.4.}}
    \label{fig:sup}
\end{figure}

\Blue{The standard defines two channel access modalities: the beacon enabled modality, which uses a slotted CSMA/CA and the optional Guaranteed Time Slot (GTS) allocation mechanism, and a simpler unslotted CSMA/CA without beacons. The communication is organized in temporal windows denoted superframes. Fig.~\ref{fig:sup} shows the superframe structure of the beacon enabled mode.}

\begin{figure}[t]
  \centering
  \subfigure[Non time-critical data packet or GTS request transmission.]
  {
      \includegraphics[width = 0.46\columnwidth]{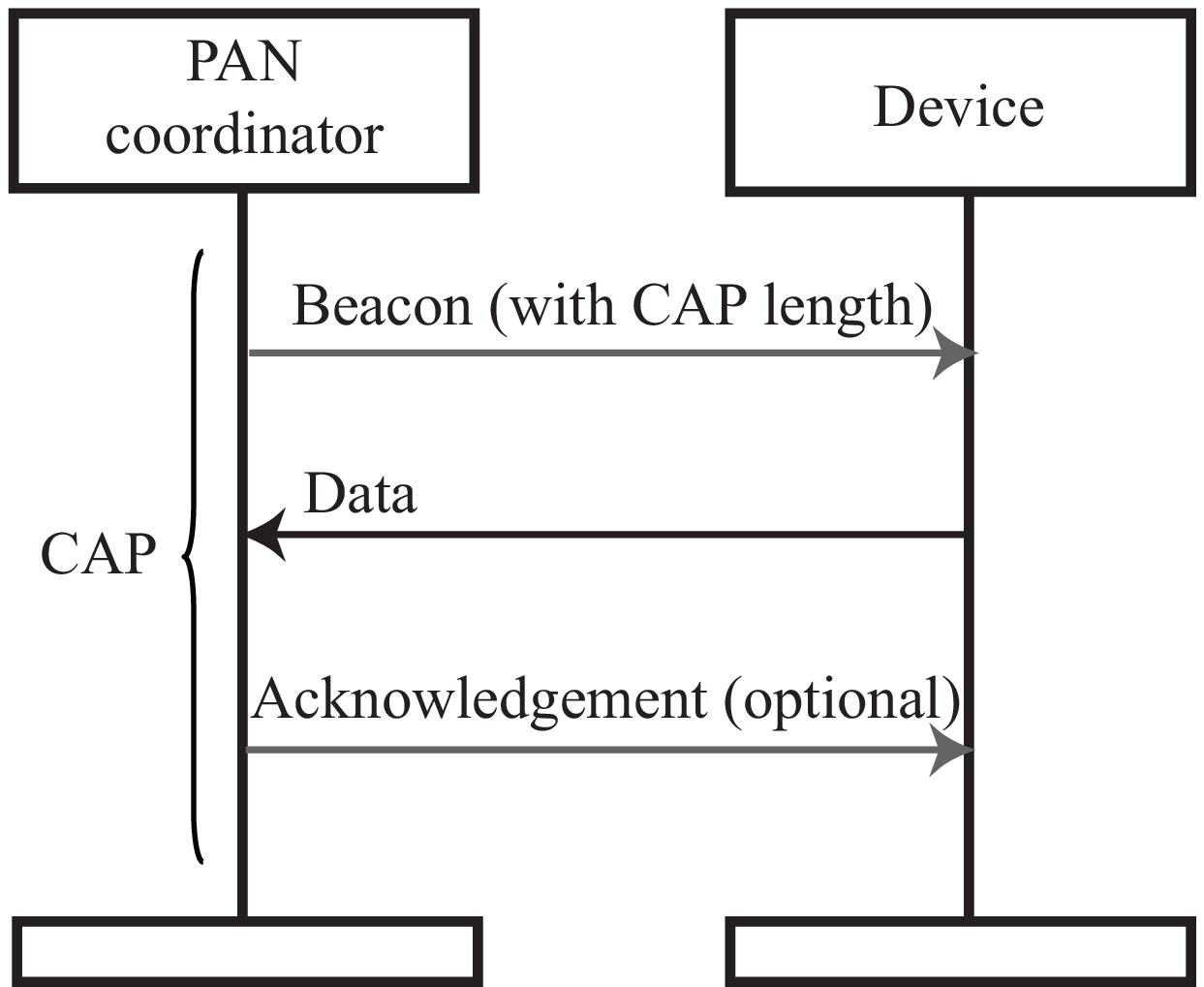}
      \label{fig:CAP_diagram}
  }
  \subfigure[Time-critical data packet transmission.]
  {
      \includegraphics[width = 0.46\columnwidth]{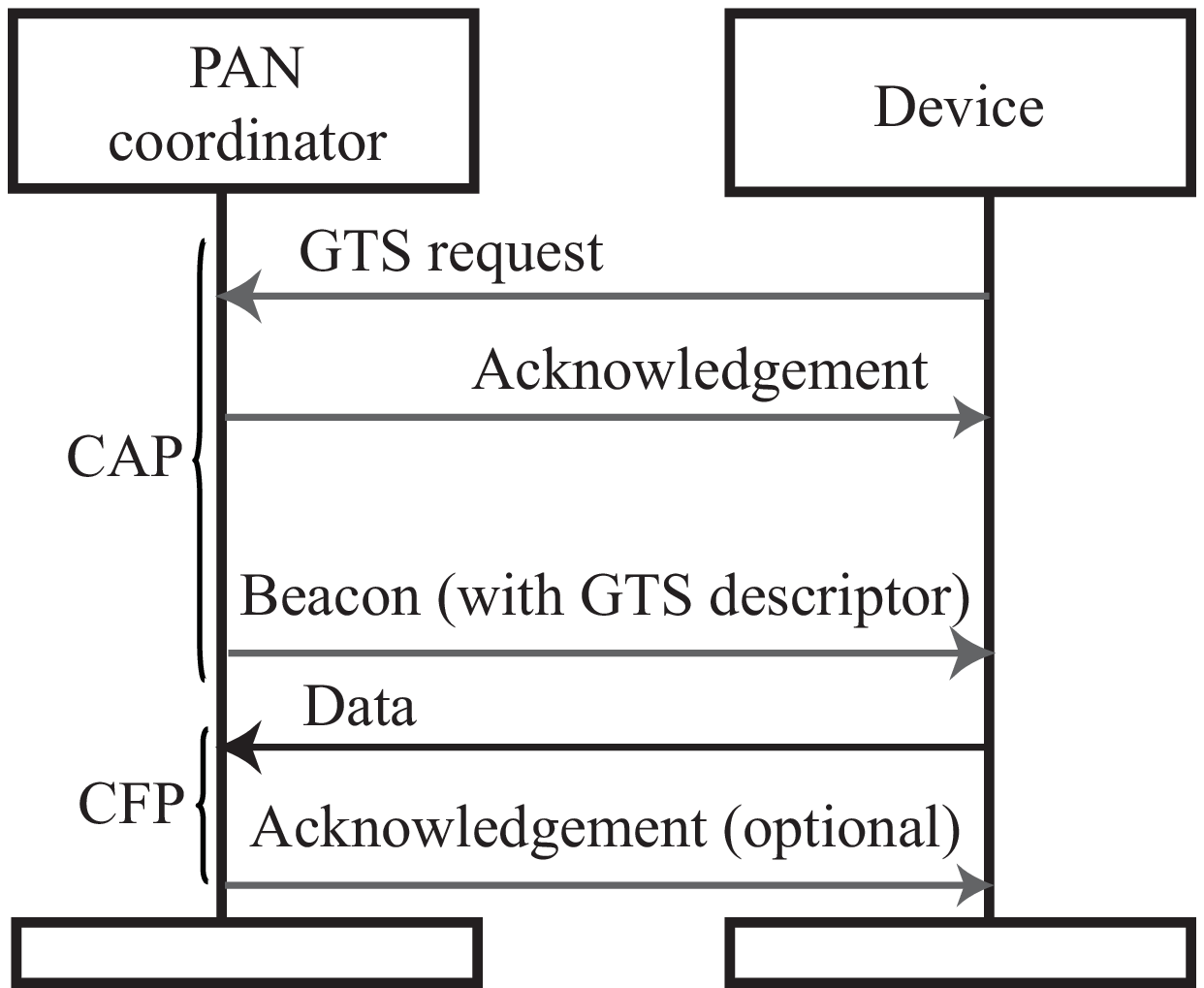}
      \label{fig:CFP_diagram}
  }
  \caption{\Blue{Data transfers of beacon enabled mode during the CAP and CFP.}}
  \label{fig:data_diagram}
\end{figure}

\Blue{In the following, we focus on the beacon enabled modality. The network coordinator periodically sends beacon frames in every beacon interval $T_{\BI}$ to identify its PAN and to synchronize nodes that communicate with it. The coordinator and nodes can communicate during the active period, called the superframe duration $T_{\SD}$, and enter the low-power mode during the inactive period. The structure of the superframe is defined by two parameters, the beacon order $(\BO)$ and the superframe order $(\SO)$, which determine the length of the superframe and its active period, given by}
\begin{align}
& T_{\BI}  =  \aBaseSuperframeDuration \times 2^{\BO} \,, \label{eq:BI}\\
& T_{\SD}  =  \aBaseSuperframeDuration \times 2^{\SO} \,,
\label{eq:SD}
\end{align}
\Blue{respectively, where $0 \leq \SO \leq \BO \leq 14$ and $\aBaseSuperframeDuration$ is the number of symbols forming a superframe when $\SO$ is equal to $0$. In addition, the superframe is divided into $16$ equally sized superframe slots of length $\aBaseSlotDuration$. Each active period can be further divided into a Contention Access Period (CAP) and an optional Contention Free Period (CFP), composed of GTSs. A slotted CSMA/CA mechanism is used to access the channel of non time-critical data frames and GTS requests during the CAP. In the CFP, the dedicated bandwidth is used for time-critical data frames. Fig.~\ref{fig:data_diagram} illustrates the date transfer mechanism of the beacon enabled mode for the CAP and CFP. In the following, we describe the data transmission mechanism for both CAP and CFP.}

\begin{figure}[]\centering
  \includegraphics[width = 0.91\columnwidth]{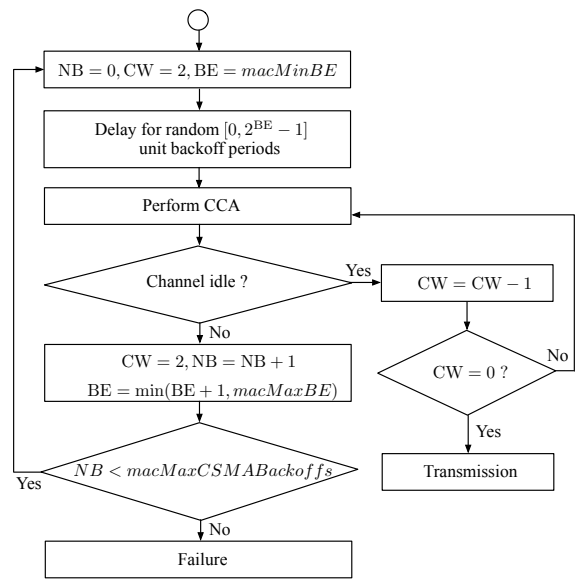}
  \caption{Slotted CSMA/CA algorithm of IEEE 802.15.4 beacon enabled mode}
  \label{fig:csma}
\end{figure}

\Blue{\textit{CSMA/CA mechanism of CAP:} CSMA/CA is used both during the CAP in beacon enabled mode and all the time in non-beacon enabled mode. In CAP, the nodes access the network by using slotted CSMA/CA  as described in Fig.~\ref{fig:csma}. The major difference of CSMA/CA in different channel access modes is that the backoff timer starts at the beginning of the next backoff slot in beacon enabled mode, and immediately in non-beacon enabled mode. Upon the request of the transmission of a packet, the following steps of the CSMA/CA algorithms are performed: 1) The channel access variables are initialized. Contention window size, denoted by $\CW$, is initialized to 2 for the slotted CSMA/CA. The backoff exponent, called $\BE$, and number of backoff stages, denoted by $\NB$, are set to $0$ and $macMinBE$, respectively. 2) A backoff time is chosen randomly from $[0, 2^{\BE}-1]$ interval. The node waits for the backoff time in units of backoff period slots. 3) When the backoff timer expires, the clear channel assessment is performed. a) If the channel is free in non-beacon enabled mode, the packet is transmitted. b) If the channel is free in beacon enabled mode, $\CW$ is updated by subtracting $1$. If $\CW=0$, the packet is transmitted. Otherwise, the second channel assessment is performed. c) If the channel is busy, the variables are updated as follows: $\NB = \NB+1, \BE = \min(\BE+1, macMaxBE), \CW=2$. The algorithm continues with step $2$ if $\NB < macMaxCSMABackoffs$, otherwise the packet is discarded.}

\textit{GTS allocation of CFP:} \Blue{The coordinator is responsible for the GTS allocation and determines the length of the CFP in a superframe. To request the allocation of a new GTS, the node sends the GTS request command to the coordinator. The coordinator confirms its receipt by sending an ACK frame within CAP. Upon receiving a GTS allocation request, the coordinator checks whether there are sufficient resources and, if possible, allocates the requested GTS. We recall that Fig.~\ref{fig:CFP_diagram} illustrates the GTS allocation mechanism. The CFP length depends on the GTS requests and the current available capacity in the superframe. If there is sufficient bandwidth in the next superframe, the coordinator determines a node list for GTS allocation based on a first-come-first-served policy. Then, the coordinator transmits the beacon including the GTS descriptor to announce the node list of the GTS allocation information. Note that on receipt of the ACK to the GTS request command, the node continues to track beacons and waits for at most \emph{aGTSDescPersistenceTime} superframes. A node uses the dedicated bandwidth to transmit the packet within the CFP.}

%

\subsubsection{WirelessHART} 
WirelessHART was released in September 2007 as the first wireless communication standard for process control applications~\cite{whart}. The standard adopts the IEEE 802.15.4 physical layer on channels 11--25 at $2.4$ GHz. TDMA is used to allow the nodes to put their radio in sleep when they are not scheduled to transmit or receive a packet for better energy efficiency and eliminate collisions for better reliability. The slot size of the TDMA is fixed at $10$ ms.

To increase the robustness to interference in the harsh industrial environments, channel hopping and channel blacklisting mechanisms are incorporated into the direct sequence spread spectrum technique adopted in the IEEE 802.15.4 standard. Frequency hopping spread spectrum is used to alternate the channel of transmission on a packet level, i.e., the channel does not change during the packet transmission. The frequency hopping pattern is not explicitly defined in the standard but needs to be determined by the network manager and distributed to the nodes. Channel blacklisting may also be used to eliminate the channels containing high interference levels. The network manager performs the blacklisting based on the quality of reception at different channels in the network.

\Blue{WirelessHART defines two primary routing approaches for multihop networks: source routing and graph routing. Source routing provides a single route of each flow, while graph routing provides multiple redundant routes~\cite{Han11}.} Since the source routing approach only establishes a fixed single path between source and destination, any link or node failure disturbs the end-to-end communication. For this reason, source routing is mostly used for network diagnostics purposes to test the end-to-end connection. \Blue{Multiple redundant routes in the graph routing provide significant improvement over source routing in terms of the routing reliability.} The routing paths are determined by the network manager based on the periodic reports received from the nodes including the historical and instantaneous quality of the wireless links.


\subsubsection{ISA-100.11a}
ISA-100.11a standard was released in September 2009 with many similar features to WirelessHART but providing more flexibility and adaptivity~\cite{isa_sp100}. Similar to WirelessHART, the standard adopts the IEEE 802.15.4 physical layer on channels 11--25 at $2.4$ GHz but with the optional additional usage of channel $26$. TDMA is again used for better energy consumption and reliability performance but with a configurable slot size on a superframe base. 

ISA-100.11a adopts channel hopping and blacklisting mechanism to improve the communication robustness similar to WirelessHART but with more flexibility. The standard adopts three channel hopping mechanisms: slotted hopping, slow hopping, and hybrid hopping. In slotted hopping, the channel is varied in each slot, same as WirelessHART. In slow hopping, the node stays on the same channel for consecutive time slots, a number which is configurable. Slow hopping facilitates the communication of nodes with imprecise synchronization, join process of new nodes, and transmission of event-driven packets. Transmissions in a slow hopping period is performed by using CSMA/CA. This mechanism decreases the delay of event-based packets while increasing energy consumption due to unscheduled transmission and reception times. In hybrid hopping, slotted hopping is combined with slow hopping by accommodating slotted hopping for periodical messages and slow hopping for less predictable new or event-driven messages. There are five predetermined channel hopping patterns in this standard, in contrast to WirelessHART that does not explicitly define hopping patterns.

\subsubsection{IEEE 802.15.4e}
\Blue{This standard has been released in 2012 with the goal of introducing new access modes to address the delay and reliability constraints of industrial applications~\cite{ieee802154e}. IEEE 802.15.4e defines three major MAC modes, namely, Time Slotted Channel Hopping (TSCH), Deterministic and Synchronous Multichannel Extension (DSME), and Low Latency Deterministic Network (LLDN). }

\Blue{\textit{Time Slotted Channel Hopping:} TSCH is a medium access protocol based on the IEEE 802.15.4 standard for industrial automation and process control~\cite{Accettura15}. The main idea of TSCH is to combine the benefits of time slotted access with multichannel and channel hopping capabilities. Time slotted access increases the network throughput by scheduling the collision-free links to meet the traffic demands of all nodes. Multichannel allows more nodes to exchange their packets at the same time by using different channel offsets. Since TSCH is based on the scheduling of TDMA slot and FDMA, the delay is deterministically bounded depending on the time-frequency pattern. Furthermore, the packet based frequency hopping is supported to achieve a high robustness against interference and other channel impairments. TSCH also supports various network topologies, including star, tree, and mesh. TSCH mode exhibits many similarities to WirelessHART and ISA-100.11a, including slotted access, multichannel communication, and frequency hopping for mesh networks. In fact, it defines more details of the MAC operation with respect to WirelessHART and ISA-100.11a.}

\Blue{In the TSCH mode, nodes synchronize on a periodic slotframe consisting of a number of time slots. Each node obtains synchronization, channel hopping, time slot and slotframe information from Enhanced Beacons (EBs) that are periodically sent by other nodes in order to advertise the network. The slots may be dedicated to one link or shared among links. A dedicated link is defined as the pairwise assignment of a directed communication between nodes in a given time slot on a given channel offset. Hence, a link between communicating nodes can be represented by a pair specifying the time slot in the slotframe and the channel offset used by the nodes in that time slot. However, the TSCH standard does not specify how to derive an appropriate link schedule.}

\Blue{Since collisions may occur in shared slots, the exponential backoff algorithm is used to retransmit the packet in the case of a transmission failure to avoid repeated collisions. Differently from the original IEEE 802.15.4 CSMA/CA algorithm, the backoff mechanism is activated only after a collision is experienced rather than waiting for a random backoff time before the transmission. }




\Blue{\textit{Deterministic and Synchronous Multichannel Extension:} DSME is designed to support stringent timeliness and reliability requirements of factory automation, home automation, smart metering, smart buildings and patient monitoring~\cite{ieee802154e}. DSME extends the beacon enabled mode of the IEEE 802.15.4 standard, relying on the superframe structure, consisting of CAPs and CFPs, by increasing the number of GTS time slots and frequency channels used~\cite{ieee802154}. The channel access of DSME relies on a specific structure called multi-superframe. Each multi-superframe consists of a collection of superframes defined in IEEE 802.15.4. The beacon transmission interval is a multiple number of multi-superframes without inactive period. By adopting a multi-superframe structure, DSME tries to support both periodic and aperiodic (or event-driven) traffic, even in large multihop networks. }


\Blue{In a DSME network, some coordinators periodically transmit an EB, used to keep all the nodes synchronized and allow new nodes to join the network. The distributed beacon and GTS scheduling algorithms of DSME allow to quickly react to time-varying traffic and changes in the network topology. Specifically, DSME allows to establish dedicated links between any two nodes of the network for the multihop mesh networks with deterministic delay. DSME is scalable and does not suffer from a single point of failure because beacon scheduling and slot allocation are performed in a distributed manner. This is the major difference with TSCH, which relies on a central entity. Given the large variety of options and features, DSME turns out to be one of the most complex modes of the IEEE 802.15.4e standard. Due to the major complexity issue, DSME still lacks a complete implementation. Moreover, all the current studies on DSME are limited to single-hop or cluster-tree networks, and do not investigate the potentialities of mesh topologies.}



\Blue{\textit{Low Latency Deterministic Network:} LLDN is designed for very low latency applications of the industrial automation where a large number of devices sense and actuate the factory production in a specific location~\cite{Ouanteur17}. Differently from TSCH and DSME, LLDN is designed only for star topologies, where a number of nodes need to periodically send data to a central sink using just one channel frequency. Specifically, the design target of LLDN is to support the data transmissions from 20 sensor nodes every 10~ms. Since the former IEEE 802.15.4 standard does not fulfill this constraint, the LLDN mode defines a fine granular deterministic TDMA access. Similarly to IEEE 802.15.4, each LLDN device can obtain the exclusive access for a time slot in the superframe to send data to the PAN coordinator. The number of time slots in a superframe determines how many nodes can access the channel. If many nodes need to send their packets, the PAN coordinator needs to equip with multiple transceivers, so as to allow simultaneous communications on different channels. }

\Blue{In LLDN, short MAC frames with just a 1-octet MAC header are used to accelerate frame processing and reduce transmission time. Moreover, a node can omit the address fields in the header, since all packets are destined to the PAN coordinator. Compared with TSCH, LLDN nodes do not need to wait after the beginning of the time slot in order to start transmitting. Moreover, LLDN provides a group ACK feature. Hence, time slots can be much shorter than the one of TSCH, since it is not necessary to accommodate waiting times and ACK frames. }

\subsubsection{6LoWPAN}
\Blue{6LoWPAN provides a compaction and fragmentation mechanism to efficiently transport IPv6 packets in IEEE 802.15.4 frames~\cite{Palattella13}. The IPv6 header is compressed by the removal of the fields that are not needed or always have the same contents, and inferring IPv6 addresses from link layer addresses. Moreover, fragmentation rules are defined so that multiple IEEE 802.15.4 frames can form one IPv6 packet. 6LoWPAN allows low-power devices to communicate by using IP.}

\subsubsection{RPL}
\Blue{RPL is an IPv6 routing protocol for Low-Power and Lossy Networks (LLNs) proposed to meet the delay, reliability and high availability requirements of critical applications in industrial and environmental monitoring~\cite{rpl_standard}. RPL is a distance vector and source routing protocol. It can operate on top of any link layer mechanism including IEEE 802.15.4 PHY and MAC. RPL adopts Destination Oriented Directed Acyclic Graphs (DODAGs), where most popular destination nodes act as the roots of the directed acyclic graphs. Directed acyclic graphs are tree-like structures that allow the nodes to associate with multiple parent nodes. The selection of the stable set of parents for each node is based on the objective function. The objective function determines the translation of routing metrics, such as delay, link quality and connectivity, into ranks, where the rank is defined as an integer, strictly decreasing in the downlink direction from the root. RPL left the routing metric open to the implementation~\cite{rpl_of0}.}


\subsubsection{6TiSCH}
\Blue{6TiSCH integrates an Internet-enabled IPv6-based upper stack, including 6LoWPAN, RPL and IEEE 802.15.4 TSCH link layer~\cite{6tisch_01}. This integration allows achieving industrial performance in terms of reliability and power consumption while providing an IP-enabled upper stack. 6TiSCH Operation Sublayer (6top) is used to manage TSCH schedule by allocating and deallocating resources within the schedule, monitor performance and collect statistics.}

\Blue{6top uses either centralized or distributed scheduling. In centralized scheduling, an entity in the network collects topology and traffic requirements of the nodes in the network, computes the schedule and then sends the schedule to the nodes in the network. In distributed scheduling, nodes communicate with each other to compute their own schedule based on the local topology information. 6top labels the scheduled cells as either hard or soft depending on their dynamic reallocation capability. A hard cell is scheduled by the centralized entity and can be moved or deleted inside the TSCH schedule only by that entity. 6top maintains statistics about the network performance in the scheduled cells. This information is then used by the centralized scheduling entity to update the schedule as needed. Moreover, this information can be used in the objective function of RPL. On the other hand, a soft cell is typically scheduled by a distributed scheduling entity. If a cell performs significantly worse than other cells scheduled to the same neighbor, it is reallocated, providing an interference avoidance mechanism in the network. The distributed scheduling policy, called on-the-fly scheduling, specifies the structure and interfaces of the scheduling~\cite{6tisch_otf}. If the outgoing packet queue of a node fills up, the on-the-fly scheduling negotiates additional time slots with the corresponding neighbors. If the queue is empty, it negotiates the removal of the time slots.}


\subsubsection{IEEE 802.11}
\Blue{The basic 802.11 MAC layer uses the Distributed Coordination Function (DCF) with a simple and flexible exponential backoff based CSMA/CA and optional RTS/CTS for medium sharing~\cite{wlan97}. If the medium is sensed idle, the transmitting node transmits its frame. Otherwise, it postpones its transmission until the medium is sensed free for a time interval equal to the sum of an Arbitration Inter-Frame Spacing (AIFS) and a random backoff interval. DCF experiences a random and unpredictable backoff delay. As a result, the periodic real-time NCS packets may miss their deadlines due to the long backoff delay, particularly under congested network conditions. }



\Blue{To enforce a timeliness behavior for WLANs, the original 802.11 MAC defines another coordination function called the Point Coordination Function (PCF). This is available only in infrastructure mode, where nodes are connected to the network through an Access Point (AP). APs send beacon frames at regular intervals. Between these beacon frames, PCF defines two periods: the Contention Free Period (CFP) and the Contention Period (CP). While DCF is used for the CP, in the CFP, the AP sends contention-free-poll packets to give them the right to send a packet. Hence, each node has an opportunity to transmit frames during the CFP. In PCF, data exchange is based on a periodically repeated cycle (e.g., superframe) within which time slots are defined and exclusively assigned to nodes for transmission. PCF does not provide differentiation between traffic types, and thus does not fulfill the deadline requirements for the real-time control systems. Furthermore, this mode is optional and is not widely implemented in WLAN devices.}

\subsubsection{IEEE 802.11e}
\Blue{As an extension of the basic DCF mechanism of 802.11, the 802.11e enhances the DCF and the PCF by using a new coordination function called the Hybrid Coordination Function (HCF)~\cite{80211e05}. Similar to those defined in the legacy 802.11 MAC, there are two methods of channel accesses, namely, Enhanced Distributed Channel Access (EDCA) and HCF Controlled Channel Access (HCCA) within the HCF. Both EDCA and HCCA define traffic categories to support various QoS requirements.}

\Blue{The IEEE 802.11e EDCA provides differentiated access to individual traffic known as Access Categories (ACs) at the MAC layer. Each node with high priority traffic basically waits a little less before it sends its packet than a node with low priority traffic. This is accomplished through the variation of CSMA/CA using a shorter AIFS and contention window range for higher priority packets. Considering the real-time requirements of NCSs, the periodic NCS traffic should be defined as an AC with a high priority~\cite{Cena10} and saturation must be avoided for high priority ACs~\cite{Sojka08}.}


\Blue{HCCA extends PCF by supporting parametric traffic and comes close to actual transmission scheduling. Both PCF and HCCA enable contention-free access to support collision-free and time-bounded transmissions. In contrast to PCF,  the HCCA allows for CFPs being initiated at almost anytime to support QoS differentiation. The coordinator drives the data exchanges at runtime according to specific rules, depending on the QoS of the traffic demands. Although HCCA is quite appealing, like PCF, HCCA is also not widely implemented in network equipment. Hence, some researches adapt the DCF and EDCA mechanisms for practical real-time control applications~\cite{Tian16, Seno17, Wei13, Heo09}.}

\subsection{Wireless Network Parameters}
\Blue{To fulfill the control system requirements, the bandwidth of the wireless networks needs to be allocated to high priority data for sensing and actuating with specific deadline requirements. However, existing QoS-enabled wireless standards do not explicitly consider the deadline requirements and thus lead to unpredictable performance of WNCS~\cite{Cena07, Tian16}. The wireless network parameters determine the probability distribution of the critical interactive system variables. Some design parameters of different layers are the transmission power and rate of the nodes, the decoding capability of the receiver at the physical layer, the protocol for channel access and energy saving mechanism at the MAC layer, and the protocol for packet forwarding at the routing layer.}

\subsubsection{Physical Layer}
The physical layer parameters that determine the values of the critical interactive system variables are the transmit power and rate of the network nodes. The decoding capability of the receiver depends on the signal-to-interference-plus-noise ratio (SINR) at the receiver and SINR criteria. SINR is obviously the ratio of the signal power to the total power of noise and interference, while SINR criteria is determined by the transmission rate and decoding capability of the receiver. The increase in the transmit power of the transmitter increases SINR at the receiver. However, the increase in the transmit power at the neighboring nodes causes a decrease at the SINR, due to the increase in interference. Optimizing the transmit power of neighboring nodes is, therefore, critical in achieving SINR requirements at the receivers. 

The transmit rate determines the SINR threshold at the receivers. As the transmit rate increases, the required SINR threshold increases. Moreover, depending on the decoding capability of the receiver, there may be multiple SINR criteria. For instance, in successive interference cancellation, multiple packets can be received simultaneously based on the extraction of multiple signals from the received composite signal, through successive decoding~\cite{interf_sic, interf_sic2}. 

IEEE 802.15.4 allows the adjustment of both transmit power and rate. However, WirelessHART and ISA-100.11a use fixed power and rate, operating at the suboptimal region. 

\subsubsection{Medium Access Control}
MAC protocols fall into one of three categories: \textit{contention-based access}, \textit{schedule-based access}, and \textit{hybrid access protocols}. 

\textit{Contention-based Access Protocol:} Contention-based random access protocols used in WNCS mostly adopt the CSMA/CA mechanism of IEEE 802.15.4. The values of the parameters that determine the probability distribution of delay, message loss probability, and energy consumption include the minimum and maximum value of backoff exponent, denoted by $macMinBE$ and $macMaxBE$, respectively, and maximum number of backoff stages, called $macMaxCSMABackoffs$. \Blue{Similarly to IEEE 802.15.4, the corresponding parameters for IEEE 802.11 MAC include the IFS time, contention window size, number of tries to sense the clean channel, and retransmission limits due to missing ACKs.}

The energy consumption of CSMA/CA has been shown to be mostly dominated by the constant listening to the channel~\cite{pedamacs, sinem_analysis}. Therefore, various energy conservation mechanisms adopting low duty-cycle operation have later been proposed~\cite{SMAC}\nocite{TMAC, BMAC}--\cite{XMAC}. In low duty-cycle operation, the nodes periodically cycle between a sleep and listening state, with the corresponding durations of sleep time and listen time, respectively. Low duty-cycle protocols may be synchronous or asynchronous. In synchronous duty-cycle protocols, the listen and sleep time of neighboring nodes are aligned in time~\cite{SMAC, TMAC}. However, this requires an extra overhead for synchronization and exchange of schedules. In asynchronous duty-cycle protocols, on the other hand, the transmitting node sends a long preamble~\cite{BMAC} or multiple short preambles~\cite{XMAC} to guarantee the wakeup of the receiver node. The duty-cycle parameters, i.e., sleep time and listen time, significantly affect the delay, message loss probability, and energy consumption of the network. Using a larger sleep time reduces the energy consumption in idle listening  at the receiver, while increasing the energy consumption at the transmitter due to the transmission of longer preamble. Moreover, the increase in sleep time significantly degrades the performance of message delay and reliability due to the high contention in the medium with increasing traffic.

\textit{Schedule-based Access Protocol:} Schedule-based protocols are based on assigning time slots, of possibly variable length, and frequency bands to a subset of nodes for concurrent transmission. Since the nodes know when to transmit or receive a packet, they can put their radio in sleep mode when they are not scheduled for any activity. The scheduling algorithms can be classified into two categories: fixed priority scheduling and dynamic priority scheduling~\cite{Liu00}. In fixed priority scheduling, each flow is assigned a fixed priority off-line as a function of its periodicity parameters, including sampling period and delay constraint. For instance, in rate monotonic and deadline monotonic scheduling, the flows are assigned priorities as a function of their sampling periods and deadlines, respectively: The shorter the sampling period and deadline, the higher the priority. Fixed priority scheduling algorithms are preferred due to their simplicity and lower scheduling overhead but are typically non-optimal since they do not take the urgency of transmissions into account. On the other hand, in dynamic priority scheduling algorithms, the priority of the flow changes over time depending on the execution of the schedule. For instance, in Earliest Deadline First (EDF) Scheduling, the transmission closest to the deadline will be given highest priority, so, scheduled next; whereas in least laxity first algorithm, the priority is assigned based on the slack time, which is defined as the amount of time left after the transmission if the transmission started now. Although dynamic priority scheduling algorithms have higher scheduling overhead, they perform much better due to the dynamic adjustment of priorities over time. 


\textit{Hybrid Access Protocol:} Hybrid protocols aim to combine the advantages of contention-based random access and schedule-based protocols: Random access eliminates the overhead of scheduling and synchronization, whereas scheduled access provides message delay and reliability guarantees by eliminating collisions. IEEE 802.15.4 already provides such a hybrid architecture for flexible usage depending on the application requirements~\cite{ieee802154}.

\subsubsection{Network Routing}
\Blue{On the network layer, the routing protocol plays an extremely important role in achieving high reliability and real-time forwarding together with energy efficiency for large scale WNCS, such as large-scale aircraft avionics and industrial automation. Various routing protocols are proposed to achieve energy efficiency for traditional WSN applications~\cite{Winter13,Zhen13}. However, to deal with much harsher and noisier environments, the routing protocol must additionally provide reliable real-time transmissions~\cite{Kumar14}. Multipath routing has been extensively studied in wireless networks for overcoming wireless errors and improving routing reliability~\cite{Ganesan01,Tarique09}. Most of previous works focus on identifying multiple link/node-disjoint paths to guarantee energy efficiency and robustness against node failures~\cite{Ganesan01, Marina01}.}

\Blue{ISA~100.11a and WirelessHART employ a simple and reliable routing mechanism called graph routing to enhance network reliability through multiple routing paths. When using graph routing, the network manager builds multiple graphs of each flow. Each graph includes some device numbers and forwarding list with unique graph identification. Based on these graphs, the manager generates the corresponding sub-routes for each node and transmits to every node. Hence, all nodes on the path to the destination are pre-configured with graph information that specifies the neighbors to which the packets may be forwarded. For example, if the link of the sub-routes is broken, then the node forwards the packet to another neighbor of other sub-routes corresponding to the same flow. There has been an increasing interest in developing new approaches for graph routing with different routing costs dependent on reliability, delay, and energy consumption~\cite{Han11,routing03,routing04}. }

\Blue{RPL employs the objective function to specify the selection of the routes in meeting the QoS requirements of the applications. Various routing metrics have been proposed in the objective function to compute the rank value of the nodes in the network. The rank represents the virtual coordinate of the node, i.e., its distance to the DODAG root with respect to a given metric. Some approaches propose the usage of a single metric, including link expected transmission count~\cite{rpl_etx1, rpl_etx2}, node remaining energy, link delay~\cite{rpl_delay}, MAC based metrics considering packet losses due to contention~\cite{rpl_mac} and queue utilization~\cite{rpl_queue1, rpl_queue2}. \Blue{\cite{rpl_combine1} proposes two methods, namely, simple combination and lexical combination, for combining two routing metrics among the hop count, expected transmission count, remaining energy, and received signal strength indicator.} In simple combination, the rank of the node is determined by using a composition function as the weighted sum of the ranks of two selected metrics. In lexical combination, the node selects the neighbor with the lower value of the first selected metric, and if they are equal in the first metric, the node selects the one with the lower value of the second composition metric. Finally,~\cite{rpl_combine2} combines a set of these metrics in order to provide a configurable routing decision depending on the application requirements based on the fuzzy parameters.}

\begin{figure}[]
  \centering
\includegraphics[width = 1\columnwidth]{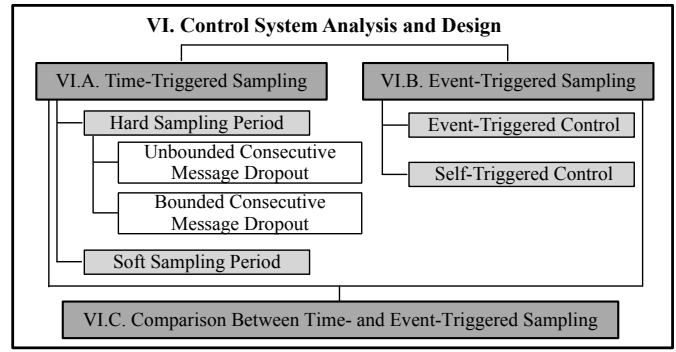} 
\caption{\Blue{Subsection structure of Section~\ref{sec:control}}} \label{fig:diagram2}
\end{figure}

\section{Control System Analysis and Design}\label{sec:control}
This section provides a brief overview of the analysis and design of control systems to deal with the non-ideal critical interactive system variables resulting from the wireless network. The presence of an imperfect wireless network degrades the performance of the control loop and can even lead to instability. Therefore, it is important to understand how these interactive system variables influence the closed-loop performance in a quantitative manner. \Blue{Fig.~\ref{fig:diagram2} illustrates the section structure and relations.}

Control system analysis has two main usages here: requirement definition for the network design and the actual control algorithm design. First, since the control cost depends on the network performance such as message loss and delay, the explicit set of requirements for the wireless network design are determined to meet a certain control performance. This allows the optimization of the network design to meet the given constraints imposed by the control system instead of just improving the reliability, delay, or energy efficiency. Second, based on the control system analysis, the controller is designed to guarantee the control performance under imperfect network operation.

Despite the interdependence between the three critical interactive variables of sampling period, message delay, and message dropout, as we have discussed in Section~\ref{sec:para}, much of the available literature on NCS considers only a subset of these variables due to the high complexity of the problem. Since any practical wireless network incurs imperfect network performance, the WNCS designers must carefully consider the performance feasibility and tradeoffs. Previous studies in the literature analyze the stability of control systems by considering either only wireless sensor--controller channel, e.g.,~\cite{Seiler05, Yue04, Seiler01} or both sensor--controller and controller--actuator, e.g.,~\cite{Branicky00, Zhang01, Xiong07, control_bounded2,  Zhang08, Zhang07,  Nilsson98, Shousong03}.


Hybrid system and Markov jump linear system have been applied for the modeling and control of NCS under message dropout and message delay. The hybrid or switched system approach refers to continuous-time dynamics with (isolated) discrete switching events~\cite{Lin09}. Mathematically, these components are usually described by a collection of indexed differential or difference equations. For NCS, a continuous-time control system can be modelled as the continuous dynamics and network effects such as message dropouts and message delays are modelled as the discrete dynamics~\cite{Xiong07, control_bounded2, Zhang08, Zhang07, Yu04}. Compared to switched systems, in Markov jump linear system the mode switches are governed by a stochastic process that is statistically independent from the state values~\cite{Costa93}. Markov systems may provide less conservative requirements than switched systems. However, the network performance must support the independent transitions between states. In other words, this technique is effective if the network performance is statistically independent or modelled as a simple Markov model.

The above theoretical approaches can be used to derive network requirements as a function of the sampling period, message dropout, and message delay. Some network requirements are explicitly related to the message dropout and message delay, such as maximum allowable message dropout probability, number of consecutive message dropouts, and message delay. Furthermore, since various analytical tools only provide sufficient conditions for closed-loop stability, their requirements might be too conservative. In fact, many existing results are shown to be conservative in simulation studies and finding tighter bounds on the network is an area of great interest~\cite{Zhang01,Yue04,Walsh02}. 

To highlight the importance of the sampling mechanism, we classify NCS analysis and design methods into \textit{time-triggered sampling} and \textit{event-triggered sampling}.

\subsection{Time-Triggered Sampling}\label{sec:time}
Time-triggered NCSs can be classified into two categories based on the relationship between sampling period and message delay: \textit{hard sampling period} and \textit{soft sampling period}. The message delay of hard sampling period is smaller than the sampling period. The network discards the message if is not successfully transmitted within its sampling period and tries to transmit the latest sampled message for the hard sampling period. On the other hand, the node of the soft sampling period continues to transmit the outdated messages even after its sampling period. The wireless network design must take into account which time-triggered sampling method is implemented.

\subsubsection{Hard Sampling Period}
The message dropouts of NCSs are generally modelled as stochastic variables with and without limited number of consecutive message dropouts. Hence, we classify hard sampling period into \textit{unbounded consecutive message dropout} and \textit{bounded consecutive message dropout}. 

\textit{Unbounded Consecutive Message Dropout:} When the controller is collocated with the actuators, a Markov jump linear system can be used to analyze the effect of the message dropout~\cite{Seiler05, Seiler01, Sinopoli04, Tatikonda04}. In~\cite{Seiler05,Seiler01}, the message dropout is modelled as a Bernoulli random process with dropout probability $p \in [0, 1)$. Under the Bernoulli dropout model, the system model of the augmented state is a special case of a discrete-time Markov jump linear system. The matrix theory is used to show exponential stability of the NCS with dropout probability $p$. The stability condition interpreted as a linear matrix inequality is a useful tool to design the output feedback controller as well as requirement derivation of the maximum allowable probability of message dropouts for the network design. However, the main results of~\cite{Seiler05, Seiler01} are hard to apply for wireless network design since they ignore the message delay for a fixed sampling period. Furthermore, the link reliability of wireless networks does not follow a Bernoulli random process since wireless links are highly correlated over time and space in practice~\cite{Beta08,Kappa10}.

While the sensor--controller communication is considered without any delays in~\cite{Seiler05, Seiler01}, the sensor--controller and controller--actuator channels are modelled as two switches indicating whether the corresponding message is dropped or not in~\cite{Zhang07}. A discrete-time switched system is used to model the closed-loop NCS with message dropouts when the message delay and sampling period are fixed. By using switched system theory, sufficient conditions for exponential stability are presented in terms of nonlinear matrix inequalities. The proposed methods provide an explicit relation between the message dropout rate and the stability of the NCS. Such a quantitative relation enables the design of a state feedback controller guaranteeing the stability of the closed-loop NCS under a certain message dropout rate. The network may assign a fixed time slot for a single packet associated to the message to guarantee the constant message delay. However, since this does not allow any retransmissions, it will significantly degrade the message dropout rate. Another way to achieve constant message delay may be to buffer the received packet at the sink. However, this will again degrade the control performance with higher average delay. 

In order to apply the results of~\cite{Seiler05, Seiler01, Zhang07}, the wireless network needs to monitor the message dropout probability and adapt its operation in order to meet the maximum allowable probability of message dropouts. These results can further be used to save network resources while preserving the stability of the NCS by dropping messages at a certain rate.



In fact, most NCS research focuses on the stability analysis and design of the control algorithm rather than explicit derivation of network requirements useful for the wireless network design. Since the joint design of controller and wireless networks necessitates the derivation of the required message dropout probability and message delay to achieve the desired control cost,~\cite{joint_01} provides the formulation of the control cost function as a function of the sampling period, message dropout probability, and message delay. Most NCS researches use the linear quadratic cost function as the control objective. The model combines the stochastic models of the message dropout~\cite{Schenato07} and the message delay~\cite{Nilsson98-thesis}. Furthermore, the estimator and controller are obtained by extending the results of the optimal stochastic estimator and controller of~\cite{Schenato07, Nilsson98-thesis}. Given a control cost, numerical methods are used to derive a set of the network requirements imposed on the sampling period, message dropout, and message delay. One of the major drawbacks is the high computation complexity to quantify the control cost in order to find the feasible region of the network requirements.



\textit{Bounded Consecutive Message Dropout:} 
Some NCS literatures~\cite{control_bounded2,Yu04} assume limited number of consecutive message dropouts, such hard requirements are unreasonable for wireless networks where the packet loss probability is greater than zero at any point in time. Hence, some other approaches~\cite{isa_sp100,Park14-tac,Karagiannis11} set stochastic constraints on the maximum allowable number of consecutive message dropouts. 

Control theory provides deterministic bounds on the maximum allowable number of consecutive message dropouts~\cite{control_bounded2, Yu04}. In~\cite{Yu04}, a switched linear system is used to model NCSs with constant message delay and arbitrary but finite message dropout over the sensor--controller channel. The message dropout is said to be arbitrary if the sampling sequence of the successfully applied actuation is an arbitrary variable within the maximum number of consecutive message dropouts. Based on the stability criterion of the switched system, a linear matrix inequality is used to analyze sufficient conditions for stability. Then, the maximum allowable bound of consecutive message dropouts and the feedback controllers are derived via the feasible solution of a linear matrix inequality. 

A Lyapunov-based characterization of stability is provided and explicit bounds on the Maximum Allowable Transfer Interval (MATI) and the Maximally Allowable Delay (MAD) are derived to guarantee the control stability of NCSs, by considering time-varying sampling period and time-varying message delays, in~\cite{control_bounded2}. If there are message dropouts for the time-triggered sampling, its effect is modelled as a time-varying sampling period from receiver point-of-view. MATI is the upper bound on the transmission interval for which stability can be guaranteed. If the network performance exceeds the given MATI or MAD, then the stability of the overall system could not be guaranteed. The developed results lead to tradeoff curves between MATI and MAD. These tradeoff curves provide effective quantitative information to the network designer when selecting the requirements to guarantee stability and a desirable level of control performance.

Many control applications, such as wireless industrial automation~\cite{isa_sp100}, air transportation systems~\cite{Park14-tac}, and autonomous vehicular systems~\cite{Karagiannis11}, set a stochastic MATI constraint in the form of keeping the time interval between subsequent state vector reports above the MATI value with a predefined probability to guarantee the stability of control systems. Stochastic MATI constraint is an efficient abstraction of the performance of the control systems since it is directly related to the deadline of the real-time scheduling of the network design~\cite{Park15_tr}.



\subsubsection{Soft Sampling Period}
Sometimes it is reasonable to relax the strict assumption on the message delay being smaller than the sampling period. Some works assume the eventual successful transmission of all messages with various types of deterministic or stochastic message delays~\cite{Branicky00,Zhang01}. Since the packet retransmission corresponding to the message is allowed beyond its sampling period, one can consider the packet loss as a message delay. While the actuating signal is updated after the message delay of each sampling period if the delay is smaller than its sampling period~\cite{Branicky00,Walsh02}, the delays longer than one sampling period may result in more than one (or none) arriving during a single sampling period. It makes the derivation of recursive formulas of the augmented matrix of closed-loop system harder, compared to the hard sampling period case.

To avoid high computation complexity, an alternative approach defines slightly different augmented state to use the stability results of switched systems in~\cite{Zhang01}. Even though the stability criterion defines the MATI and MAD requirements, there are fundamental limits of this approach to apply for wireless networks. The stability results hold if there is no message dropout for the fixed sampling period and constant message delay, since the augmented matrix consiered is a function of the fixed sampling period with the constant message delay. Hence, the MATI and MAD requirements are only used to set the fixed sampling period and message delay deadline. On the other hand, the NCS of~\cite{control_bounded2} uses the time-varying sampling and varying message delay to take into account the message dropout and stochastic message delay. Hence, the MATI and MAD requirements of~\cite{control_bounded2} are more practical control constraints than the ones of~\cite{Zhang01} to apply to wireless network design.





In~\cite{Shousong03}, a stochastic optimal controller is proposed to compensate long message delays of the sensor--controller channel for fixed sampling period. The stochastic delay is assumed to be bounded with a known probability density function. Hence, the network manager needs to provide the stochastic delay model by analyzing delay measurements. In both~\cite{Zhang01} and~\cite{Shousong03}, the NCSs assume the eventual successful transmission of all messages. This approach is only reasonable if MATI is large enough compared to the sampling period to guarantee the eventual successful transmission of messages with high probability. However, it is not applicable for fast dynamical system (i.e., small MATI requirement). 



While~\cite{Zhang01, Shousong03} do not explicitly consider message dropouts,~\cite{Yue04} jointly considers the message dropout and message delay longer than the fixed sampling period over the sensor--controller channel. From the derived stability criteria, the controller is designed and the MAD requirement is determined under a fixed message dropout rate by solving a set of matrix inequalities. Even though the message dropout and message delay are considered, the tradeoff between performance measures is not explicitly derived. However, it is still possible to obtain tradeoff curves by using numerical methods. The network is allowed to transmit the packet associated to the message within the MAD. The network also monitors the message dropout rate. Stability is guaranteed if the message dropout rate is lower than its maximum allowable rate. Furthermore, the network may discard outdated messages to efficiently utilize the network resource as long as the message dropout rate requirement is satisfied.

\subsection{Event-Triggered Sampling}
Event-triggered control is reactive since it generates sensor measurements and control commands when the plant state deviates more than a certain threshold from a desired value. On the other hand, self-triggered control is proactive since it computes the next sampling or actuation instance ahead of current time. Event- and self-triggered control have been demonstrated to significantly reduce the network traffic load~\cite{self-event,Arzen99}. Motivated by those advantages, a systematic design of event-based implementations of stabilizing feedback control laws was performed in~\cite{Tabuada07}.

Event-triggered and self-triggered control systems consist of two elements, namely, a feedback controller that computes the control command, and a triggering mechanism that determines when the control input has to be updated again. The triggering mechanism directly affects the traffic load~\cite{Arzen99}. There are many proposals for the triggering rule in the event-triggered literature. Suppose that the state $x(t)$ of the physical plant is available. One of the traditional objectives of event-triggered control is to maintain the condition
\begin{align}
\parallel x(t) - x(t_{k}) \parallel \leq \delta,
\end{align}
where $t_{k}$ denotes the time instant when the last control task is executed (the last event time) and $\delta > 0$ is a threshold~\cite{Lunze10}. The next event time instant is defined as 
\begin{align}
t_{k+1} = \inf \left \{ t > t_{k} | \parallel x(t) - x(t_{k}) \parallel > \delta \right \} \,. 
\end{align}

The sensor of the event-triggered control loop continuously monitors the current plant state and evaluates the triggering condition. Network traffic is generated if the plant state deviates by the threshold. The network design problem is particularly challenging because the wireless network must support the randomly generated traffic. Furthermore, event-triggered control does not provide high energy efficiency since the node must continuously activate the sensing part of the hardware platform.

Self-triggered control determines its next execution time based on the previously received data and the triggering rule~\cite{Wang09}. Self-triggered control is basically an emulation of an event-triggered rule, where one considers the model of the plant and controller to compute the next triggering time. Hence, it is predictive sampling based on the plant models and controller rules. This approach is generally more conservative than the event-triggered approach since it is based on approximate models and predicted events. The explicit allocation of network resources based on these predictions improves the real-time performance and energy efficiency of the wireless network. However, since event- and self-triggered control generate fewer messages, the message loss and message delay might seem to be more critical than for time-triggered control~\cite{self-event}.

\subsection{Comparison Between Time- and Event-Triggered Sampling} 
\Blue{One of the fundamental issues is to compare the performance of time-triggered sampling and event-triggered sampling approaches by using various channel access mechanisms~\cite{Cervin08, Blind11_1, Blind11_2}. In fact, many event-based control researches show performance improvement since it often reduces the network utilization~\cite{Jose14, Cervin08}. However, recent works of the event-based control using the random access show control performance limitations in the case when there are a large number of control loops~\cite{Blind11_1, Blind11_2}. \cite{Cervin08} considers a control system where a number of time-triggered or event-triggered control loops are closed over a shared communication network. This research is one of the inspiring works of WNCS co-design problem, where both the control policy and network scheduling policy have been taken into account. The overall target of the framework is to minimize the sum of the stationary state variance of the control loops. A Dirac pulse is applied to achieve the minimum plant state variance as the control law. The sampling can be either time-triggered or event-triggered, depending on the MAC schemes such as the traditional TDMA, FDMA, and CSMA schemes. Intuitively, TDMA is used for the time-triggered sampling, while the event-triggered sampling is applied for CSMA. Based on the previous work~\cite{Johannesson07}, the event-triggered approach is also used for FDMA since the event-triggered sampling with a minimum event interval $T$ performs better than the one using the time-triggered sampling with the same time interval $T$. The authors of~\cite{Cervin08} assume that once the MAC protocol gains the network resource, the network is busy for specific delay from sensor to actuator, after which the control command is applied to the plant. The simulation results show that event-triggered control using CSMA gives the best performance. Even though the main tradeoffs and conclusions of the paper are interesting, some assumptions are not realistic. In practice, the Dirac pulse controls are unrealistic due to the capability limit of actuators. For simplicity, the authors assume that the contention resolution time of CSMA is negligible compared to the transmission time. This assumption is not realistic for general wireless channel access schemes such as IEEE 802.15.4 and IEEE 802.11. Furthermore, the total bandwidth resource of FDMA is assumed to scale in proportion to the number of plants, such that the transmission delay from sensor to actuator is inversely proportional to the number of plants. These assumptions are not practical since the frequency spectrum is a limited resource for general wireless networks, thus further studies are needed.}

\Blue{While most previous works on event-based control consider a single control loop or small number of control loops, \cite{Blind11_1} compares time-triggered control and event-based control for a NCS consisting of a large number of plants. The pure ALOHA protocol is used for the event-based control of NCSs. The authors show that packet losses due to collisions drastically reduce the performance of event-based control if packets are transmitted whenever the event-based control generates an event. Remark that the instability of the ALOHA network itself is a well known problem in communications~\cite{Rom90}. It turns out that in this setup time-triggered control is superior to event-based control. The same authors also analyze the tradeoff between delay and loss for event-based control with slotted ALOHA~\cite{Blind11_2}. They show that the slotted ALOHA significantly improve the control cost of the state variance respect to the one of the pure ALOHA. However, the time-triggered control still performs better. Therefore, it is hard to generalize the performance comparison between time-triggered sampling and event-triggered sampling approaches since it really depends on the network protocol and topology.}

\section{Wireless Network Design Techniques for Control Systems}\label{sec:design}
\begin{figure}[]
  \centering
\includegraphics[width = 1\columnwidth]{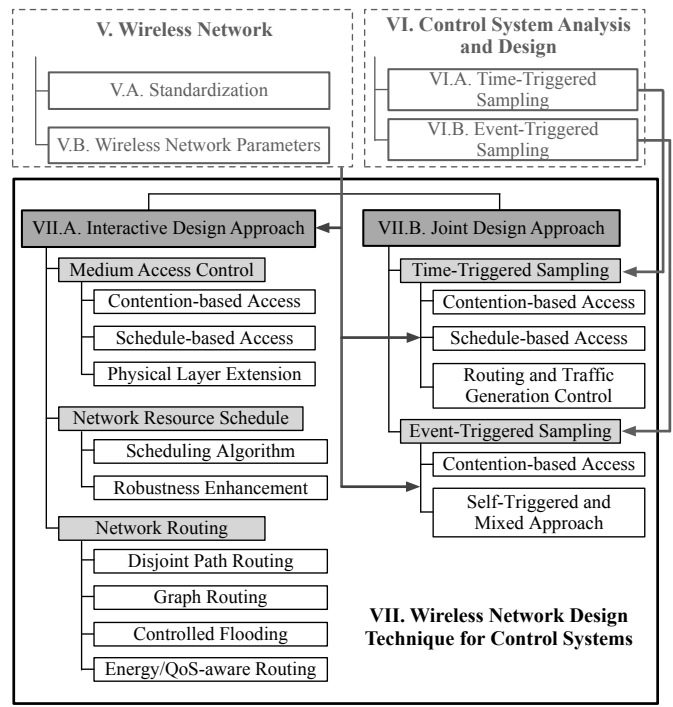} 
\caption{\Blue{Subsection structure of Section~\ref{sec:design} related to previous Sections~\ref{sec:net} and~\ref{sec:control}}} \label{fig:diagram3}
\end{figure}

This section presents various design and optimization techniques of wireless networks for WNCS. We distinguish \textit{interactive design approach} and \textit{joint design approach}. In the interactive design approach, the wireless network parameters are tuned to satisfy given constraints on the critical interactive system variables, possibly enforced by the required control system performance. In the joint design approach, the wireless network and control system parameters are jointly optimized considering their interaction through the critical system variables. \Blue{Fig.~\ref{fig:diagram3} illustrates the section structure related to previous Sections~\ref{sec:net} and~\ref{sec:control}. In Table~\ref{tab:comp}, we summarize the characteristics of the related works. In the table, we have demonstrated whether indications of requirements and communication and control parameters have been included in the network design or optimization for WNCS. Table~\ref{tab:class} classifies previous design approaches of WNCS based on control and communication aspects. Furthermore, Table~\ref{tab:stadard_class} categorizes previous works based on the wireless standards described in Section~\ref{sec:standard}.}

\subsection{Interactive Design Approach}

In the interactive design approach, wireless network parameters are tuned to satisfy the given requirements of the control system. Most of the interactive design approaches assume time-triggered control systems, in which sensor samples are generated periodically at predetermined rates. They generally assume that the requirements of the control systems are given in the form of upper bounds on the message delay or message dropout with a fixed sampling period. \Blue{The adoption of wireless communication technologies for supporting control applications heavily depends on the ability to guarantee the bounded service times for messages, at least from a probabilistic point of view. This aspect is particularly important in control systems, where the real-time requirement is considered much more significant than other performance metrics, such as throughput, that are usually important in other application areas. Note that the real-time performance of wireless networks heavily depends on the message delay and message dropout. Hence, we mainly discuss the deadline-constrained MAC protocols of IEEE 802.15.4 and IEEE 802.11. Different analytical techniques can provide the explicit requirements of control systems for wireless networks, as we discussed in Section~\ref{sec:control}. The focus of previous research is mainly on the design and optimization of MAC, network resource scheduling, and routing layer, with limited efforts additionally considering physical layer parameters.}



\begin{table*}[h]
\scriptsize
\centering
\caption{\Blue{Comparison of related works. The circle with plus $\bigoplus$ denotes that the paper explicitly considers the indication of the column. The dot $\bigcirc$ denotes that the paper does not include the indication and hence cannot control it, but simulation or experiment results include it. The terms ``the'', ``sim'', ``exp'' of evaluation column mean that the proposed solution is evaluated through theoretical analysis, simulation, or realistic experiment, respectively. }} \label{tab:comp}
\begin{tabular}{| C{1.3cm} | C{0.3cm} | C{0.4cm} | C{0.8cm}  | C{0.7cm} | C{0.6cm} || C{0.5cm} | C{0.4cm} | C{0.8cm} | C{1.0cm} | C{0.7cm}  | C{0.8cm} | C{1.2cm} || C{0.8cm} || C{1.0cm} |}
\hline
{} & \multicolumn{5}{c||}{Requirements} & \multicolumn{7}{c||}{System Parameters} & \multirow{3}{*}{Scenarios} & \multirow{3}{*}{Evaluation}   \\ \cline{7-13}
{} & \multicolumn{5}{c||}{} & \multicolumn{5}{c|}{Communication Parameters} & \multicolumn{2}{c||}{Control Parameters} & {} & {} \\ \cline{2-13}
{} & Loss & Delay & Sampling Period & Control Cost & Energy & Power & Rate & Schedule & Contention & Routing & Sampling Period & Control Algorithm & {} & {} \\ \hline  
\cite{Tian16}  & $\bigoplus$ & $\bigoplus$  & {}  & {} & {} &  {} & {} & {} & $\bigoplus$ & {}  & {}  & {} & 1-hop & the/sim \\ \hline
\cite{sinempan_exp}  & $\bigoplus$ & $\bigoplus$  & {}  & {} & $\bigoplus$ &  {} & {} & {} & $\bigoplus$ & {}  & {}  & {} & 1-hop & the/exp \\ \hline
\cite{Cena07}  & $\bigcirc$ & $\bigcirc$  & {}  & {} & {} &  {} & {} & {} & $\bigoplus$ & {}  & {}  & {} & 1-hop & the/exp \\ \hline
\cite{random_06}  & $\bigcirc$ & $\bigoplus$  & {}  & {} & {} &  {} & {} & {} & $\bigoplus$ & {}  & {}  & {} & 1-hop & the/sim \\ \hline
\cite{random_07}  & $\bigoplus$ & {}  & {}  & {} & $\bigoplus$ &  {} & {} & {} & $\bigoplus$ & $\bigoplus$  & {}  & {} & multihop & the/sim \\ \hline
\cite{Seno17}  & $\bigcirc$ & $\bigoplus$  & $\bigcirc$  & {} & {} &  {} & {} & $\bigoplus$ & $\bigcirc$ & {}  & {}  & {} & 1-hop & the/sim/exp \\ \hline
\cite{Wei13}  & $\bigcirc$ & $\bigcirc$  & $\bigoplus$  & {} & {} &  {} & {} & $\bigoplus$ & {} & {}  & {}  & {} & 1-hop & the/sim/exp \\ \hline
\cite{Moraes08-fcs}  & $\bigcirc$ & $\bigcirc$  & $\bigcirc$  & {} & {} &  {} & {} & $\bigoplus$ & {} & {}  & {}  & {} & 1-hop & the \\ \hline
\cite{Toscano11}  & $\bigcirc$ & $\bigcirc$  & $\bigcirc$  & {} & {} &  {} & {} & $\bigoplus$ & {} & {}  & {}  & {} & 1-hop & sim \\ \hline
\cite{Boggia08}  & $\bigcirc$ & $\bigcirc$  & $\bigcirc$  & {} & {} &  {} & {} & $\bigoplus$ & {} & {}  & {}  & {} & 1-hop & exp \\ \hline
\cite{sinem_tdma}  & {} & $\bigcirc$  & {}  & {} & {} &  {} & {} & $\bigoplus$ & {} & {}  & {}  & {} & multihop & the/sim \\ \hline
\cite{Yan14}  & $\bigcirc$ & $\bigoplus$  & {}  & {} & {} &  {} & {} & $\bigoplus$ & {} & {}  & {}  & {} & multihop & the/sim \\ \hline
\cite{separate_08, separate_09}  & {} & $\bigoplus$  & {}  & {} & {} &  {} & {} & $\bigoplus$ & {} & {}  & {}  & {} & multihop & the/sim \\ \hline
\cite{Saifullah15}  & $\bigcirc$ & $\bigoplus$  & {}  & {} & {} &  {} & {} & $\bigoplus$ & {} & {}  & {}  & {} & multihop & the/sim/exp \\ \hline
\cite{separate_11}  & $\bigcirc$ & $\bigcirc$  & {}  & {} & {} &  {} & $\bigcirc$ & $\bigcirc$ & {} & {}  & {}  & {} & multihop & the/exp \\ \hline
\cite{separate_16}  & $\bigoplus$ & $\bigoplus$  & {}  & {} & {} &  {} & {} & $\bigoplus$ & {} & {}  & {}  & {} & 1-hop & the/sim \\ \hline
\cite{separate_14}  & $\bigoplus$ & $\bigoplus$  & {}  & {} & {} &  {} & {} & $\bigoplus$ & {} & {}  & {}  & {} & multihop & the/exp \\ \hline
\cite{separate_15}  & $\bigoplus$ & $\bigoplus$  & {}  & {} & {} &  {} & {} & $\bigoplus$ & {} & {}  & {}  & {} & 1-hop & the/sim/exp \\ \hline
\cite{sinem_twc}  & $\bigcirc$ & $\bigoplus$  & $\bigoplus$  & {} & $\bigoplus$ &  $\bigoplus$ & $\bigoplus$ & $\bigoplus$ & {} & {}  & {}  & {} & 1-hop & the/sim \\ \hline
\cite{Saifullah10} & {} & $\bigoplus$ & {}  & {} & {}  &  {} & {} & $\bigoplus$  & {} & {}  & {}  & {} & multihop & the/sim/exp \\ \hline
\cite{Zhibo14}  &  $\bigcirc$ & $\bigoplus$  & {}  & {} & {} &  {} & {} & {} & {} & $\bigoplus$  & {}  & {} & multihop & the/sim \\ \hline
\cite{routing03, routing04}  & $\bigcirc$ & $\bigcirc$  & {}  & {} & {} &  {} & {} & {} & {} & $\bigoplus$  & {}  & {} & multihop & the/sim \\ \hline
\cite{Ganesan01}  & $\bigcirc$ & {}  & {}  & {} & $\bigoplus$ &  {} & {} & {} & {} & $\bigoplus$  & {}  & {} & multihop & the/sim \\ \hline
\cite{Sadi13}  & $\bigcirc$ & $\bigoplus$ & $\bigoplus$  & {} & $\bigcirc$ &  $\bigoplus$ & $\bigoplus$ & $\bigoplus$ & {} & {}  & $\bigoplus$  & {} & 1-hop & the/sim \\ \hline
\cite{Saifullah14_rate}  & $\bigcirc$ & $\bigoplus$ & {}  & $\bigoplus$ & {}  &  {} & {} & {}  & {} & {}  & $\bigoplus$  & $\bigcirc$ & multihop & the/sim \\ \hline
\cite{Breath}  & $\bigoplus$ & $\bigoplus$ & {}  & {} & $\bigoplus$  &  $\bigoplus$ & {} & {}  & $\bigoplus$ & $\bigoplus$  & {}  & {} & multihop & the/exp \\ \hline
\cite{Park16_cloc}  &  $\bigcirc$ & $\bigcirc$  & $\bigcirc$  & $\bigcirc$ & {} &  {} & $\bigoplus$ & {} & {} & $\bigoplus$  & $\bigoplus$  & {} & multihop & the/sim/exp \\ \hline
\cite{Sadi14}  &  $\bigcirc$ & $\bigcirc$  & $\bigcirc$  & $\bigcirc$ & $\bigoplus$ &  {} & $\bigoplus$ & $\bigoplus$ & {} & {}  & $\bigoplus$  & {} & 1-hop & the/sim \\ \hline
\cite{sinem_icc}  &  $\bigcirc$ & $\bigcirc$  & $\bigcirc$  & $\bigcirc$ & $\bigoplus$ &  {} & $\bigoplus$ & $\bigoplus$ & {} & {}  & $\bigoplus$  & {} & 1-hop & the/sim \\ \hline
\cite{burak_opt}  &  $\bigoplus$ & $\bigoplus$  & {}  & $\bigoplus$ & $\bigoplus$ &  {} & {} & $\bigoplus$ & {} & {}  & $\bigoplus$  & $\bigoplus$ & multihop & the/sim \\ \hline
\cite{Alur11}  &  $\bigoplus$ & $\bigoplus$  & {}  & $\bigoplus$ & {} &  {} & {} & $\bigoplus$ & {} & $\bigoplus$  & {}  & $\bigoplus$ & multihop & the/sim \\ \hline
\cite{Ding13, Wang14}  &  $\bigcirc$ & $\bigcirc$  & {}  & $\bigoplus$ & {} &  {} & {} & $\bigoplus$ & {} & {}  & {}  & $\bigoplus$ & 1-hop & the/sim/exp \\ \hline
\cite{joint_06, Bai10, Ulusoy11}  &  $\bigoplus$ & $\bigoplus$  & {}  & $\bigoplus$ & {} &  {} & {} & {} & $\bigoplus$ & {}  & {}  & $\bigoplus$ & 1-hop & the/sim \\ \hline
\cite{Li16}  & $\bigcirc$ & {} & {}  &  $\bigoplus$ & $\bigcirc$  &  {} & {} & $\bigcirc$  & {} &  $\bigoplus$  & $\bigcirc$  & $\bigoplus$ & multihop & sim/exp \\ \hline
\cite{Li09}  & $\bigcirc$ & $\bigoplus$ & $\bigcirc$  & {} & $\bigcirc$  &  {} & {} & {}  & {} & $\bigoplus$  & {}  & {} & multihop & sim \\ \hline
\cite{Park12-adaptive}  &  $\bigoplus$  & $\bigoplus$  & {}  & {} & $\bigoplus$   &  {} & {} & {}  & $\bigoplus$  & {}  & {}  & {} & 1-hop & the/sim/exp\\ \hline
\cite{Dobslaw15}  & $\bigoplus$ & $\bigcirc$ & $\bigcirc$  & {} & {}  &  {} & {} & $\bigoplus$  & {} & {}  & {}  & {} & multihop & the/sim \\ \hline
\cite{Dong15}  & $\bigcirc$ & $\bigcirc$ & {}  & {} & {}  &  {} & {} & $\bigoplus$  & $\bigoplus$ & $\bigcirc$  & {}  & {} & multihop & the/sim/exp \\ \hline
\cite{Han11}  & $\bigcirc$ & {} & {}  & {} & {}  &  {} & {} &  $\bigoplus$  & {} &  $\bigoplus$  & {}  & {} & multihop & sim \\ \hline
\cite{routing02}  & $\bigoplus$ &  $\bigoplus$ & {}  & {} &  $\bigoplus$  &  {} & {} & {}  & {} &  $\bigoplus$  & {}  & {} & multihop & the/exp \\ \hline
\cite{Heo09}  & $\bigoplus$ & $\bigoplus$ & {}  & {} & $\bigoplus$  &  {} & {} & {}  & {} & $\bigoplus$  & {}  & {} & multihop & sim \\ \hline
\cite{Quang12} & {} & $\bigoplus$ & {}  & {} & $\bigcirc$  &  {} & {} & {}  & {} & $\bigoplus$  & {}  & {} & multihop & sim \\ \hline
\cite{Jose14} &  {} & {} & {}  & $\bigoplus$ & $\bigcirc$  &  {} & {} & $\bigcirc$  & {} & {}  & $\bigoplus$  & $\bigoplus$ & 1-hop & the/exp \\ \hline
\cite{joint_03}  & {} & {} & {}  & $\bigoplus$ & {}  &  {} & $\bigcirc$ & {}  & $\bigcirc$ & {}  &  $\bigcirc$ & $\bigcirc$ & 1-hop & sim \\ \hline
\cite{joint_02}  & $\bigoplus$ & $\bigoplus$ & {}  & $\bigoplus$ & {}  &  {} & {} & $\bigoplus$  & {} & {}  & $\bigoplus$  & $\bigcirc$ & multihop & the/sim \\ \hline
\cite{joint_01}  & $\bigoplus$ & $\bigoplus$ & {}  & $\bigoplus$ & $\bigoplus$  &  {} & {} & {}  & $\bigoplus$ & {}  & $\bigoplus$  & $\bigoplus$ & 1-hop & the/sim \\ \hline
\cite{Maben09}  &  {} & {} & {}  & $\bigoplus$ & {}  &  {} & {} & {}  & $\bigoplus$ & {}  & $\bigoplus$  & $\bigoplus$ & 1-hop & the \\ \hline
\cite{Cao13}  &  $\bigoplus$ & $\bigcirc$  & {}  & $\bigoplus$ & $\bigoplus$ &  {} & {} & $\bigcirc$ & {} & {}  & $\bigoplus$  & $\bigoplus$ & 1-hop & the/sim \\ \hline
\cite{Henriksson15}  &  $\bigcirc$ & $\bigcirc$  & {}  & $\bigoplus$ & {} &  {} & {} & $\bigoplus$ & {} & {}  & $\bigoplus$  & $\bigoplus$ & 1-hop & the/sim \\ \hline
\cite{Peng16}  &  {} & $\bigcirc$  & {}  & $\bigoplus$ & $\bigcirc$ &  {} & {} & $\bigcirc$ & {} & {}  & $\bigoplus$  & $\bigoplus$ & 1-hop & the/sim/exp \\ \hline
\cite{Vilgelm16, Gatsis16, Ramesh16, Mamduhi14, Molin14}  &  $\bigoplus$ & {}  & {}  & $\bigoplus$ & {} &  {} & {} & {} & $\bigoplus$ & {}  & $\bigoplus$  & $\bigoplus$ & 1-hop & the/sim \\ \hline
\cite{demirel13}  &  $\bigoplus$ & $\bigoplus$  & {}  & $\bigoplus$ & {} &  {} & {} & {} & $\bigoplus$ & {}  & $\bigoplus$  & $\bigoplus$ & 1-hop & the/sim \\ \hline
\end{tabular}
\end{table*}



\begin{table*}[t]
\scriptsize
\centering
\caption{\Blue{Classification of WNCS design techniques}} \label{tab:class}
\begin{tabular}{| C{2.8cm} | C{2.9cm} || C{3.2cm} | C{3.2cm}  | C{3.2cm} |}
\hline
\multicolumn{2}{|c||} {} & \multirow{2}{*}{Interactive Design} & \multicolumn{2}{c|}{Joint Design Approach} \\  \cline{4-5}
\multicolumn{2}{|c||} {} & {} & \multirow{1}{*}{Time-Triggered Sampling} & \multicolumn{1}{c|}{Event-Triggered Sampling} \\ \hline
\multirow{3}{*}{Medium Access Control} & Contention-based Access & {\cite{Ko06, Ai04, random_07, random_06, Merlin10, Cena07, sinem_analysis, Park12-adaptive, sinempan_analytical, sinempan_exp, Tian16}} & {\cite{joint_01, joint_03, joint_06, Bai10, Kottenstette08, Ulusoy11, Gokturk08}}  & {\cite{Maben09, demirel13, Vilgelm16, Gatsis16, Ramesh16, Mamduhi14, Molin14}} \\ \cline{2-5}
{} & Schedule-based Access & {\cite{Seno17, Wei13, Moraes08-fcs, Toscano11, Boggia08}} & {\cite{Sadi14, sinem_icc, joint_02, joint_03, Ding13, Wang14}}  & {\cite{Jose14, self-control-carlo, Cao13, Henriksson15, Peng16}} \\ \cline{2-5}
{} & Physical Layer Extension & {\cite{Sadi13, sinem_twc, separate_11, separate11_ref19, separate11_ref20, Tramarin16}} & {-}  & {-} \\ \hline
\multirow{2}{*}{Network Resource Schedule} & Scheduling Algorithm & {\cite{sinem_tdma, Dobslaw15, Yan14, separate_08, Saifullah15, Saifullah10, separate_09}} & {\cite{Alur11, burak_opt, Park16_cloc}}  & {-} \\ \cline{2-5}
{} & Robustness Enhancement & {\cite{Dong15,separate_16, separate_14, separate_15}} & {-}  & {-} \\ \hline
\multirow{4}{*}{Network Routing} & Disjoint Path & {\cite{Ganesan01, Marina01}} & {-}  & {-} \\ \cline{2-5}
{} & Graph & {\cite{routing03, routing04, Han11, routing02}} & {\cite{Li16}}  & {-} \\ \cline{2-5}
{} & Controlled Flooding & {\cite{Zhibo14, Yu12}} & {-}  & {-} \\ \cline{2-5}
{} & Energy/QoS-aware & {\cite{Breath, Heo09, Li09, Quang12}} & {-}  & {-} \\ \hline
\multicolumn{2}{|c||}{Traffic Generation Control} & {-} & {\cite{Saifullah14_rate, Park15_tr}}  & {-} \\ \hline
\end{tabular}
\end{table*}

\begin{table*}[t]
\scriptsize
\centering
\caption{\Blue{Classification of WNCS design techniques based on the wireless standards}} \label{tab:stadard_class}
\begin{tabular}{| C{2.8cm} | C{2.9cm} || C{3.2cm} | C{3.2cm}  | C{3.2cm} |}
\hline
\multicolumn{2}{|c||} {} & \multirow{2}{*}{Interactive Design} & \multicolumn{2}{c|}{Joint Design Approach} \\  \cline{4-5}
\multicolumn{2}{|c||} {} & {} & \multicolumn{1}{c|}{Time-Triggered Sampling} & \multicolumn{1}{c|}{Event-Triggered Sampling} \\ 
\hline
\multirow{3}{*}{802.15.4} {} & Physical Layer &  {\cite{separate_11, separate11_ref19, separate11_ref20}} & {-}  & {-}  \\ \cline{2-5}
& Contention & {\cite{Ko06, Ai04, random_07, random_06, sinem_analysis, Park12-adaptive, separate_15}} & {\cite{joint_01}}  & {-} \\ \cline{2-5}
{} & Hybrid &  {-} & {-}  & {\cite{Jose14, self-control-carlo, Peng16}}  \\ \hline
\multirow{2}{*}{WirelessHART}  & Schedule &  {\cite{Dobslaw15, Yan14, Saifullah10, Dong15, separate_08, Saifullah15, separate_09} } & {\cite{Alur11, burak_opt, Park16_cloc}}  & {-} \\ \cline{2-5}
{} & Routing &  {\cite{routing03, routing04, Han11, routing02, Zhibo14, Yu12}} & {\cite{Li16}}  & {-}  \\ \hline
\multirow{2}{*}{802.11} & Contention &  {\cite{Tian16, Heo09}} & {\cite{Bai10, Ulusoy11}}  & {-}  \\ \cline{2-5}
{} & Schedule &  {\cite{Seno17, Wei13, Moraes08-fcs, Toscano11, Boggia08, separate_16}} & {\cite{Ding13, Wang14}}  & {-}  \\ \hline
\multicolumn{2}{|c||}{802.11 e/g/n} &  {\cite{Cena07, Tramarin16}} & {-}  & {-}  \\ \hline
\end{tabular}
\end{table*}

\subsubsection{Medium Access Control}
\Blue{Research on real-time 802.15.4 and 802.11 networks can be classified into two groups. The first group of solutions called \textit{contention-based access} includes adaptive MAC protocols for QoS differentiations. They adapt the parameters of backoff mechanism, retransmissions, and duty-cycling dependent on the constraints. The second group called \textit{schedule-based access} relies on the contention free scheduling of a single-hop netowrk.}\\
\textit{Contention-based Access:} Contention-based random access protocols for WNCS aim to tune the parameters of the CSMA/CA mechanism of IEEE 802.15.4 and IEEE 802.11, and duty-cycling to improve delay, packet loss probability, and energy consumption performance. The adaptive tuning algorithms are either measurement-based or model-based adaptation.

The measurement-based adaptation techniques do not require any network model but rather depend on the local measurements of packet delivery characteristics. Early works of IEEE 802.15.4 propose adaptive algorithms to dynamically change the value of only a single parameter. \cite{Ko06, Ai04} adaptively determine minimum contention window size, denoted by $macMinBE$, to decrease the delay and packet loss probability of nodes and increase overall throughput. The references~\cite{random_07, random_06} extend these studies to autonomously adjust all the CSMA/CA parameters. The ADAPT protocol~\cite{random_07} adapts the parameter values with the goal of minimizing energy consumption while meeting packet delivery probability based on their local estimates. However, ADAPT tends to oscillate between two or more parameters sets. This results in high energy consumption. \cite{random_06} solves this oscillation problem by triggering the adaptation mechanism only upon the detection of a change in operating conditions. Furthermore,~\cite{Merlin10} aims to optimize duty-cycle parameters based on a linear increase/linear decrease of the duty-cycle depending on the comparison of the successfully received packet rate and its target value while minimizing the energy consumption. 

Model-based parameter optimization mainly use theoretical or experimental-based derivations of the probability distribution of delay, packet error probability, and energy consumption. A Markov model per node of IEEE 802.15.4 is used~\cite{sinem_analysis} to capture the state of each node at each moment in time. These individual Markov chains are then coupled by the memory introduced by fixed duration two slot clear channel assessment. The proposed Markov model is used to derive an analytical formulation of both throughput and energy consumption in such networks. The extension of this work in~\cite{Park12-adaptive} leads to the derivation of the reliability, delay, and energy consumption as a function of all the CSMA/CA protocol parameters for IEEE 802.15.4. 

The paper~\cite{sinempan_analytical} provides analytical models of delay, reliability, and energy consumption as a function of duty-cycle parameters by considering their effects on random backoff of IEEE 802.15.4 before successful transmissions. These models are then used to minimize energy consumption given constraints on delay and reliability. On the other hand,~\cite{sinempan_exp} derives experimental based models by using curve fitting techniques and validation through extensive experiments. An adaptive algorithm was also proposed to adjust the coefficients of these models by introducing a learning phase without any explicit information about data traffic, network topology, and MAC parameters.

\Blue{By considering IEEE 802.11, a deadline-constrained MAC protocol with QoS differentiation is presented for soft real-time NCSs~\cite{Tian16}. It handles periodic traffic by using two specific mechanisms, namely, a contention-sensitive backoff mechanism and a deadline-sensitive retry limit assignment mechanism. The backoff algorithm offers bounded backoff delays, whereas the deadline-sensitive retry limit assignment mechanism differentiates the retry limits for periodic traffic in terms of their respective deadline requirements. A Markov chain model is established to describe the proposed MAC protocol and evaluate its performance in terms of throughput, delay, and reliability under the critical real-time traffic condition. }

\Blue{\cite{Cena07} provides experimental measures and the analysis of 802.11g/e network to better understand the statistical distribution of delay for real-time industrial applications. The statistical distribution of network delay is first evaluated experimentally when the traffic patterns they support resemble the realistic industrial scenarios under the varying background traffic. Then, experimental results have been validated by means of a theoretical analysis for unsaturated traffic condition, which is a quite common condition in well-designed industrial communication systems. The performance evaluation shows that delays are generally bounded if the traffic on the industrial WLAN is light (below 20\%). If the traffic grows higher (up to 40\%), the QoS mechanism provided by EDCA is used to achieve quasi-predictable behavior and bounded delays for selected high priority messages.}



\Blue{\textit{Schedule-based Access:} The explicit scheduling of transmissions allows to meet the strict delay and reliability constraints of the nodes, by giving priority to the nodes with tighter constraint. To support soft real-time industrial applications,~\cite{Seno17} combines a number of various mechanisms of IEEE 802.11 such as transmission and retransmission scheduling, seamless channel redundancy, and basic bandwidth management to improve the deterministic network performance. The proposed protocol relies on centralized transmission scheduling of a coordinator according to the EDF strategy. Furthermore, the coordinator takes care of the number of retransmissions to achieve both delay and reliability over lossy links. In addition to scheduling, the seamless channel redundancy concurrently transmits the copies of each frame on multiple distinct radio channels. This mechanism is appealing for real-time systems since it improves the reliability without affecting timeliness. Moreover, the bandwidth manager reallocates the unused bandwidth of failed data transmission to additional attempts of other data transmissions within their deadlines.}


\Blue{\cite{Wei13} presents the design and implementation of a real-time wireless communication protocol called RT-WiFi to support high-speed control systems which typically require 1KHz or higher sampling rate. RT-WiFi is a TDMA data link layer protocol based on IEEE 802.11 physical layer. It provides deterministic timing performance on packet delivery. Since different control applications have different communication requirements on data delivery, RT-WiFi provides a configurable platform to adjust the design tradeoffs including sampling rate, delay variance, and reliability.}


\Blue{The middleware proposed in~\cite{Moraes08-fcs} uses a TDMA-based method on top of 802.11 CSMA to assign specific time slots to each real-time node to send its traffic. In~\cite{Toscano11}, a polling-based scheduling using the EDF policy on top of 802.11 MAC is incorporated with a feedback mechanism to adjust the maximum number of transmission attempts. Moreover,~\cite{Boggia08} implements a real-time communication architecture based on the 802.11 standard and on the real-time networking framework RTnet~\cite{Kiszka05}. Wireless Ralink RT2500 chipset of RTnet is used to support the strict network scheduling requirements of real-time systems. The performance indicators such as packet loss ratio and delay are experimentally evaluated by varying protocol parameters for a star topology. Experimental results show that a proper tuning of system parameters can support robust real-time network performance.}

\begin{figure}[]
  \center
  \psfrag{x}[][]{\footnotesize{(a) Robust schedule}}
  \psfrag{y}[][]{\footnotesize{ (b) EDF schedule}}
  \includegraphics[width = 1\columnwidth]{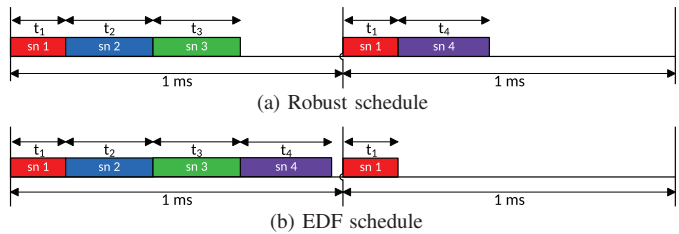}  
  \caption{\Blue{Illustrative example of two schedulers} }
   \label{fig:objective}
\end{figure}

\textit{Physical Layer Extension:} \Blue{\cite{Sadi13, sinem_twc} propose a priority assignment and scheduling algorithm as a function of sampling periods and transmission deadlines to provide maximum level of adaptivity, to accommodate the packet losses of time-triggered nodes and the transmissions of event-triggered nodes. The adaptivity metric is illustrated using the following example. Let us assume that the network consists of $4$ sensor nodes, denoted by sensor node $i$ for $i \in [1, 4]$. The packet generation period and transmission time of sensor $1$ are $1$ ms and $t_1=0.15$ ms, respectively. The packet generation period of sensor nodes $2$, $3$ and $4$ is $2$ ms, whereas packet transmission times are given by $t_2=0.20$ ms, $t_3=0.25$ ms and $t_4=0.30$ ms, respectively. Figs.~\ref{fig:objective}(a) and~\ref{fig:objective}(b) show a robust schedule where the time slots are uniformly distributed over time and the EDF schedule, respectively. The schedule given in Fig.~\ref{fig:objective}(a), is more robust to packet losses than the EDF schedule given in Fig.~\ref{fig:objective}(b). Indeed, suppose that the data packet of sensor $1$ in the first $1$ ms is not successfully transmitted. In Fig.~\ref{fig:objective}(a), the robust schedule includes enough unallocated intervals for the retransmission of sensor $1$, whereas the EDF schedule does not. Furthermore, the robust scheduler can accommodate event-triggered traffic with smaller delay than the EDF schedule, as shown in Fig.~\ref{fig:objective}. To witness, suppose that an additional packet of $0.2$ ms transmission time is generated by an event-triggered sensor node at the beginning of the scheduling frame. Then the event-triggered packet transmission can be allocated with a delay of $0.60$ ms in the robust schedule and $1.15$ ms in the EDF schedule.}

\begin{figure}[]
  \center
  \psfrag{x}[][]{\footnotesize{Number of nodes}}
  \psfrag{y}[][]{\footnotesize{Maximum delay}}
   \includegraphics[width = 0.82\columnwidth]{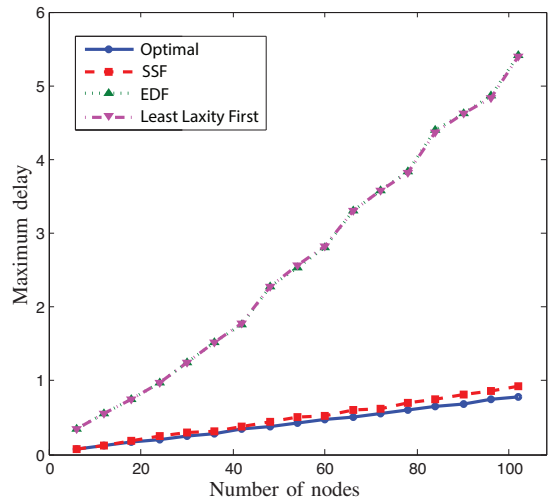}
  \caption{\Blue{Comparison of the maximum delay experienced by event-triggered components for SSF, EDF, least laxity first, and optimal scheduling algorithms.}} 
  \label{fig:max_delay}
\end{figure}

\Blue{This uniform distribution paradigm is quantified as minimizing the maximum total active length of all subframes, where the subframe length is the minimum packet generation period among the components and the total active length of a subframe is the sum of the transmission time of the components allocated to that subframe. The proposed Smallest Period into the Shortest Subframe First (SSF) algorithm has been demonstrated to significantly decrease the maximum delay experienced by the packet of an event-triggered component compared to the EDF schedule, as shown in Fig.~\ref{fig:max_delay}. Moreover, when time diversity, in the form of the retransmission of the lost packets, is included in this framework, the proposed adaptive framework decreases the average number of missed deadlines per unit time, which is defined as the average number of packets that cannot be successfully transmitted within their delay constraint, significantly compared to the EDF schedule.}

\Blue{Since IEEE 802.11n encompasses several enhancements at both PHY and MAC layers of WLAN,~\cite{Tramarin16} analyzes the performance indicators such as service time and reliability of IEEE 802.11n for industrial communication systems. The authors present both theoretical analysis and its validation through a set of experiments. The experimental analysis shows the possibility to select the IEEE 802.11n parameters to ensure the deterministic behavior for the real-time applications. In particular, it is shown that a good MIMO configuration of the standard enhances the communication reliability while sacrificing the network throughput.}


\subsubsection{Network Resource Schedule}  
\Blue{Several scheduling algorithms are proposed to efficiently assign the time slot  and the channel of the multihop networks in order to meet the strict delay and reliability requirements.}  

\textit{Scheduling Algorithm:} Some scheduling algorithms focus on meeting a common deadline for all the packets generated within a sampling period~\cite{sinem_tdma, Dobslaw15, Yan14}. \cite{sinem_tdma} formulates the delay minimization of the packet transmissions from the sensor nodes to the common access point. The optimization problem has been shown to be NP-hard. The proposed scheduling algorithms provide upper bounds on the packet delivery time, by considering many-to-one transmission characteristics. The formulation and scheduling algorithms, however, do not take packet losses into account. \cite{Dobslaw15, Yan14} introduce novel procedures to provide reliability in case of packet failures. \cite{Yan14} proposes an optimal schedule increment strategy based on the repetition of the most suitable slot until the common deadline. The objective of the optimization problem is to maximize end-to-end reliability while providing end-to-end transmission delay guarantees. The physical network nodes have been reorganized into logical nodes for improved scheduling flexibility. Two scheduling algorithms have been evaluated: dedicated scheduling and shared scheduling. In dedicated scheduling, the packets are only transmitted in the scheduled time slots, whereas in shared scheduling, the packets share scheduled time slots for better reliability. \cite{Dobslaw15} proposes a faster scheduling algorithm for the same problem introduced in~\cite{Yan14}. The algorithm is based on gradually increasing a network model from one to multiple transmitted packets as a function of given link qualities to guarantee end-to-end reliability. These scheduling algorithms can be combined with multiple path routing algorithms. The authors assume  Bernoulli distribution for the arrival success of the packets over each link. Moreover, they do not consider the transmission power, rate and packet length as a variable, assigning exactly one time slot to each transmission.

The scheduling algorithms that consider the variation of sampling periods and deadlines of the nodes over the network fall into one of two categories: fixed priority and dynamic priority. The end-to-end delay analysis of periodic real-time flows from sensors to actuators in a WirelessHART network under fixed priority scheduling policy has been performed in~\cite{separate_08}. The upper bound on the end-to-end delay of the periodic flows is obtained by mapping their scheduling to real-time multi-processor scheduling and then exploiting the response time analysis of the scheduling. Both the channel contention and transmission conflict delay due to higher priority flows have been considered. Channel contention happens when all channels are assigned to higher priority flows in a transmission slot, whereas transmission conflict occurs when there exists a common node with a transmission of higher priority flow. This study has later been extended for reliable graph routing to handle transmission failures through retransmissions and route diversity in~\cite{Saifullah15}. Similarly, both worst-case and probabilisitic delay bounds have been derived by considering channel contention and transmission conflicts. These analyses consider multihop multichannel networks with fixed time slots without incorporating any transmit power or rate adjustment mechanism.

The real-time dynamic priority scheduling of periodic deadline-constrained flows in a WirelessHART network has been shown to be NP-hard in~\cite{Saifullah10}. Upon determining necessary condition for schedulability, an optimal branch-and-bound scheduling is proposed, effectively discarding infeasible branches in the search space. Moreover, a faster heuristic conflict-aware least laxity first algorithm is developed by assigning priorities to the nodes based on the criticality of their transmission. The conflict-aware laxity is defined as the laxity after discarding time slots that can be wasted while waiting to avoid transmission conflicts. The lower the conflict-aware laxity, the higher the transmission criticality. The algorithm does not provide any guarantee on the timely packet delivery. \cite{separate_09} provides the end-to-end delay analysis of periodic real-time flows from sensors to actuators under EDF policy. The delay is bounded by considering the channel contention and transmission conflict delays. The EDF has been shown to outperform fixed priority scheduling in terms of real-time performance. 


\textit{Robustness Enhancement:} The predetermined nature of schedule-based transmissions allows the incorporation of various retransmission mechanisms in case of packet losses at random time instants. Although explicit scheduling is used to prevent various types of conflict and contention, still transmission failures may occur due to multipath fading and external interference in harsh and unstable environments. Some of the retransmission mechanisms have been introduced at the link layer~\cite{separate_11, separate11_ref19, separate11_ref20}. Since schedule is known apriori by the nodes in the network, the retransmissions can be minimized by exploiting the determinism in the packet headers to recover the unknown bytes of the header~\cite{separate_11}. Moreover, various efficient retransmission procedures can be used to minimize the number of bits in the retransmissions~\cite{separate11_ref19, separate11_ref20}. \cite{separate11_ref19} uses symbol decoding confidence, whereas~\cite{separate11_ref20} uses received signal strength variations to determine the parts of the packet received in error, so, should be retransmitted.

The retransmission mechanisms at the network layer aim to determine the best timing and quantity of shared and/or separate time slots given the link quality statistics~\cite{Dong15,separate_16, separate_14, separate_15}. \cite{separate_16} combines the retransmissions with real-time worst-case scheduling analysis. The number of possible retransmissions of a packet is limited considering the corresponding deadline and already guaranteed delay bounds of other packets. \cite{separate_14} proposes a scheduling algorithm that provides delay guarantees for the periodic real-time flows considering both link bursts and interference. A new metric called maximum burst length is defined as the maximum length of error burst, estimated by using empirical data. The algorithm then provides reliability guarantee by allocating each link one plus its corresponding maximum burst length time slots. A novel least-burst-route algorithm is used in conjunction with this scheduling algorithm to minimize the sum of worst case burst lengths over all links in the route. Similarly,~\cite{separate_15} increases the spacing between the actual transmission and the first retransmission for maximum reliability instead of allocating all the time slots in between. \cite{Dong15} improves the retransmission efficiency by using limited number of shared slots efficiently through fast slot competition and segmented slot assignment. Shared resources are allocated for retransmission due to its unpredictability. Fast slot competition is introduced by embedding more than one clear channel assessment at the beginning of the shared slots to reduce the rate of collision. On the other hand, segmented slot assignment provides the retransmission chances for a routing hop before its following hop arrives.

\subsubsection{Network Routing}
There has been increasing interest in developing efficient multipath routing to improve the network reliability and energy efficiency of wireless networks. Previous works of the multipath routings are classified into four categories based on the underlying key ideas of the routing metric and the operation: \textit{disjoint path routing}, \textit{graph routing}, \textit{controlled flooding}, and \textit{energy/QoS-aware routing}.

\textit{Disjoint Path Routing:} Most of previous works focus on identifying multiple disjoint paths from source to destination to guarantee the routing reliability against node or link failures since multiple paths may fail independently~\cite{Ganesan01, Marina01}. The disjoint paths have two types: node-disjoint and link-disjoint. While node-disjoint paths do not have any relay node in common, link-disjoint paths do not have any common link but may have common nodes. \cite{Ganesan01} provides the node-disjoint and braided multipath schemes to provide the resilience against node failures. Ad-hoc On-demand Multipath Distance Vector (AOMDV) is a multipath extension of a well-studied single path routing protocol of Ad-hoc On-demand Distance Vector (AODV)~\cite{Marina01}. 

\textit{Graph Routing:} Graph routing of ISA~100.11a and WirelessHART leads to significant improvement over a single path in terms of worst-case reliability due to the usage of multiple paths. Since the standards do not explicitly define the mechanism to build these multiple paths, it is possible to use the existing algorithms of the disjoint path. Multiple routing paths from each node to the destination are formed by generating the subgraphs containing all the shortest paths for each source and destination pair~\cite{routing03}. Real-time link quality estimation is integrated into the generation of subgraphs for better reliability in~\cite{routing04}. \cite{Han11} proposes an algorithm to construct three types of reliable routing graphs, namely, uplink graph, downlink graph, and broadcast graph for different communication purposes. While the uplink graph is a graph that connects all nodes upward to the gateway, the downlink graph of the gateway is a graph to send unicast messages to each node of the network. The broadcast graph connects gateway to all nodes of the network for the transmission of operational control commands. Three algorithms are proposed to build these graphs based on the concepts of $(k, m)$-reliability where $k$ and $m$ are the minimum required number of incoming and outgoing edges of all nodes excluding the gateway, respectively. The communication schedule is constructed based on the traffic load requirements and the hop sequence of the routing paths.

Recently, the graph routing problem has been formulated as an optimization problem where the objective function is to maximize network lifetime, namely, the time interval before the first node exhausts its battery, for a given connectivity graph and battery capacity of nodes~\cite{routing02}. This optimization problem has been shown to be NP-hard. A suboptimal algorithm based on integer programming and a greedy heuristic algorithm have been proposed for the optimization problem. The proposed algorithm shows significant improvement in the network lifetime while guaranteeing the high reliability of graph routing.



\textit{Controlled Flooding:} Previous approaches of disjoint routing and graph routing focus on how to build the routing paths and distribute the traffic load over the network. Some control applications may define more stringent requirements on the routing reliability in the harsher and noisier environments. To address the major reliability issue, a reliable Real-time Flooding-based Routing protocol (REALFLOW) is proposed for industrial applications~\cite{Zhibo14}. REALFLOW controls the flooding mechanism to further improve the multipath diversity while reducing the overhead. Each node transmits the received packet to the corresponding multiple routing paths instead of all feasible outgoing links. Furthermore, it discards the duplicated packets and outdated packets to reduce the overhead. For both uplink and downlink transmissions, the same packets are forwarded according to the related node lists in all relay nodes. Due to redundant paths and flooding mechanism, REALFLOW can be tolerant to network topology changes. Furthermore, since related node lists are distributively generated, the workloads of the gateway are greatly reduced. The flooding schedule is also extended by using the received signal strength in~\cite{Yu12}.


\textit{Energy/QoS-aware Routing:} Even though some multipath routings such as disjoint path, graph routing, and controlled flooding lead to significant reliability improvement, they also increase the cost of the energy consumption. Energy/QoS-aware routing jointly considers the application requirements and energy consumption of the network~\cite{Pan11-thesis}. Several energy-balanced routing strategies are proposed to maximize the network lifetime while meeting the strict requirements for industrial applications. 

Breath is proposed to ensure a desired packet delivery and delay probabilities while minimizing the energy consumption of the network~\cite{Breath}. The protocol is based on randomized routing, MAC, and duty-cycling jointly optimized for energy efficiency. The design approach relies on a constrained optimization problem, whereby the objective function is the energy consumption and the constraints are the packet reliability and delay. The optimal working point of the protocol is achieved by a simple algorithm, which adapts to traffic variations and channel conditions with negligible overhead. 

EARQ is another energy aware routing protocol for reliable and real-time communications for industrial applications~\cite{Heo09}. EARQ is a proactive routing protocol, which maintains an ongoing routing table updated through the exchange of beacon messages among neighboring nodes. A beacon message contains expected values such as energy cost, residual energy of a node, reliability and end-to-end message delay. Once a node gets a new path to the destination, it will broadcast a beacon message to its neighbors. When a node wants to send a packet to the destination, next hop selections are based on the estimations of energy consumption, reliability, and deadlines. If the packet chooses a path with low reliability, the source will forward a redundant packet via other paths.

\cite{Ibrahim08} proposes the minimum transmission power cooperative routing algorithm, reducing the energy consumption of a single route while guaranteeing certain throughput. However, the algorithm ignores the residual energy and communication load of neighboring nodes, which result in unbalanced energy consumption among nodes. In addition, in~\cite{Dai09}, a load-balanced routing algorithm is proposed where each node always chooses the next-hop based on the communication load of neighboring nodes. However, the algorithm has heavy computation complexity and the communication load is high. \cite{Li09, Quang12} propose a two-hop information-based routing protocol, aiming at enhancing real-time performance with energy efficiency. The routing decision in~\cite{Li09} is based on the integration of the velocity information of two-hop neighbors with energy balancing mechanism, whereas the routing decision in~\cite{Quang12} is based on the number of hops from source to destination and two-hop information of the velocity.

\subsection{Joint Design Approach}

In the joint design approach, the wireless network and control system parameters are jointly optimized considering the tradeoff between their performances. These parameters include the sampling period for time-triggered control and level crossings for event-triggered control in the control system, and transmission power and rate at the physical layer, the access parameters and algorithm of the MAC protocol, duty-cycle parameters, and routing paths in the communication system. The high complexity of the problem led to different abstractions of control and communication systems, many of which considering only a subset of these parameters. 

\subsubsection{Time-Triggered Sampling}


The joint design approaches of the time-triggered control are classified into three categories based on the communication layers: \textit{contention-based access}, \textit{schedule-based access}, and \textit{routing and traffic generation control}. 

\textit{Contention-based Access:} The usage of contention-based protocols in the joint optimization of control and communication systems requires modeling the probabilistic distribution of delay and packet loss probability in the wireless network and its effect on the control system~\cite{joint_01, joint_03, joint_06}. A general framework for the optimization of the sampling period together with link layer parameters has been first proposed in~\cite{joint_03}. The objective of the optimization problem is to maximize control system performance given the delay distribution and the packet error probability constraints. The linear quadratic cost function is used as the control performance measure. Simplified models of packet loss and delay are assumed for the contention-based random access mechanism without considering spatial reuse. The solution strategy is based on an iterative numerical method due to the complexity of the control cost used as an objective function of the optimization problem. \cite{joint_06} aims to minimize the mean-square error of the state estimation subject to delay and packet loss probability induced by the contention-based random access. The mean-square error of the estimator is derived as a function of sampling period and delay distribution under the Bernoulli random process of the packet losses.

\begin{figure}[]
    \centering
    \psfrag{X}[][]{\scriptsize{Sampling period (s)}}
    \psfrag{A}[][]{\scriptsize{Control cost}}
    \psfrag{B}[][]{\scriptsize{Throughput}}
    \psfrag{D}[][]{\scriptsize{$J^{\mathrm{i}}_{\infty}$}}
    \psfrag{E}[][]{\scriptsize{$J^{\mathrm{r}}_{\infty}$}}
    \psfrag{F}[][]{\scriptsize{Throughput}}        
    \psfrag{C}[][]{\scriptsize{$J_{\mathrm{req}}$}}
    \psfrag{S}[][]{\scriptsize{$\mathcal{S}$}}
    \psfrag{L}[][]{\scriptsize{$\mathcal{L}$}}
    \includegraphics[width = 0.9\columnwidth]{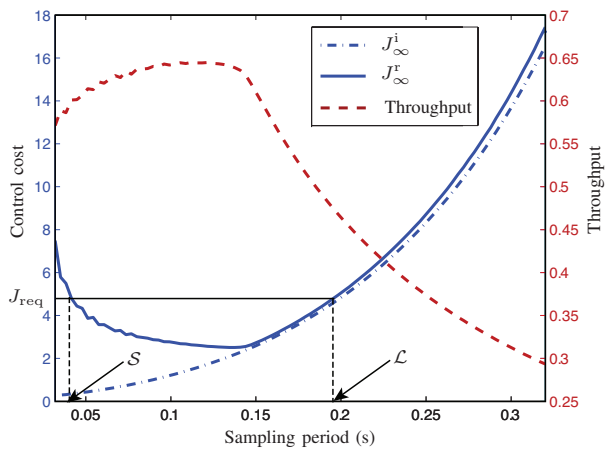}
    \caption{\Blue{Quadratic control cost of control systems and throughput of wireless networks over different sampling periods. $J^{\mathrm{i}}_{\infty}$ and $J^{\mathrm{r}}_{\infty}$ refer to the control cost bound by using the ideal network and the realistic 802.15.4 network, respectively.}} \label{fig:tradeoff}
\end{figure}

\Blue{\cite{joint_01} discusses several fundamental tradeoffs of WNCS over IEEE 802.15.4 networks. Fig.~\ref{fig:tradeoff} shows the quadratic control cost and communication throughput over different sampling periods. In the figure, $J^{\mathrm{i}}_{\infty}$ and $J^{\mathrm{r}}_{\infty}$ refer to the control cost bound using an ideal network (no packet loss and no delay) and a realistic lossy network of IEEE 802.15.4, respectively. Due to the absence of packet delays and losses, the control performance using an ideal network increases monotonically as the sampling period increases. However, when using a realistic network, a shorter sampling period does not minimize the control cost, because of the higher packet loss probability and delay when the traffic load is high. In addition, the two curves of the control cost $J^{i}_{\infty}$ and $J^{\mathrm{r}}_{\infty}$ coincide for longer sampling periods, meaning that when the sampling period is larger, the sampling period is the dominant factor in the control cost compared to the packet loss probability and delay.}

\Blue{In Fig.~\ref{fig:tradeoff}, if we consider a desired maximum control cost $J_{\mathrm{req}}$ greater than the minimum value of the control cost, then we have the feasible range of the sampling periods between $\mathcal{S}$ and $\mathcal{L}$. However, the performance of the wireless network is still heavily affected by the operating point of the sampling period. Let us consider two feasible sampling periods $\mathcal{S}$ and $\mathcal{L}$. By choosing $\mathcal{L}$, the throughput of the network is stabilized (cf.,~\cite{Rom90}), the control cost is also stabilized with respect to small perturbations of the network operation. Furthermore, the longer sampling period $\mathcal{L}$ leads to lower network energy consumption than the one of the shorter sampling period $\mathcal{S}$. Based on these observations, an adaptation of the WNCS is proposed by considering a constrained optimization problem.} The objective is to minimize the total energy consumption of the network subject to a desired control cost. The variables of the problem include both sampling period and MAC parameters of IEEE 802.15.4. The network manager predicts the energy consumption corresponding to each feasible network requirement. The optimal network requirements are obtained to minimize the energy consumption of the network out of the feasible set of network requirements.




\Blue{\cite{Bai10} proposes an interesting approach to the design of WNCS by decomposing the overall concerns into two design spaces. In the control layer, a passive control structure of~\cite{Kottenstette08} is used to guarantee the stability of NCSs. The overall NCS performance is then optimized by adjusting the retransmission limits of the IEEE 802.11 standard. At the control layer, the authors leverage their passivity-based architecture to handle the message delay and message loss. The authors consider a passive controller which produces a trajectory of the plant to track and define the control performance as its absolute tracking error. Through extensive simulation results, a convex relationship between the retransmission limit of IEEE 802.11 and the control performance is shown. Based on this observation, a MAC parameter controller is introduced to dynamically adjust the retransmission limit to track the optimal tradeoff between packet losses and delays and thus to optimize the overall control system performance. Simulation results show that the MAC adaptation can converge to a proper retransmission limit which optimizes the performance of the control system. Even though the proposed approach is interesting, the fundamental tradeoff relationships between communication parameters and control performance are not trivial to derive in practice.}

\Blue{\cite{Ulusoy11} presents a MPC-based NCS and its implementation over wireless relay networks of IEEE 802.11 and cooperative MAC protocol~\cite{Gokturk08}. The proposed approach deals with the problem from the control perspective. It basically employs a MPC, an actuator state, and an adaptive IEEE 802.11 MAC to reduce unbounded packet delay and improve the tolerance against the packet loss. Furthermore, the cooperative MAC protocol~\cite{Gokturk08} is used to improve the control performance by enabling reliable and timely data transmission under harsh wireless channel conditions.}

\textit{Schedule-based Access:} A novel framework for the communication--control joint optimization is proposed encompassing efficient abstraction of control system in the form of stochastic MATI and MAD constraints~\cite{Sadi14, sinem_icc, sinem_twc}. We should remember that MATI and MAD are defined as the maximum allowed time interval between subsequent state vector reports and the maximum allowed packet delay for the transmission, respectively, as we have discussed in Section~\ref{sec:control}. Since such hard real-time guarantees cannot be satisfied by a wireless network with non-zero packet loss probability, stochastic MATI is introduced with the goal of keeping the time interval between subsequent state vector reports above the MATI value with a predefined probability to guarantee the stability of control systems. \Blue{Further, a novel schedulability constraint in the form of forcing an adaptive upper bound on the sum of the utilization of the nodes, defined as the ratio of their delay to their sampling periods, is included to guarantee the schedulability of transmission under variable transmission rate and sampling period values. The objective of the optimization is to minimize the total energy consumption of the network while guaranteeing MATI and MAD requirements of the control system and maximum transmit power and schedulability constraints of the wireless communication system. The solution for the specific case of M-ary quadrature amplitude modulation and EDF scheduling is based on the reduction of the resulting mixed-integer programming problem into an integer programming problem based on the analysis of the optimality conditions, and relaxation of this reduced problem~\cite{Sadi14}. The formulation is also extended for any non-decreasing function of the power consumption of the nodes as the objective, any modulation scheme, and any scheduling algorithm in~\cite{sinem_icc, sinem_twc}. First, an exact solution method based on the analysis of the optimality conditions and smart enumeration techniques is introduced. Then, two polynomial-time heuristic algorithms adopting intelligent search space reduction and smart searching techniques are proposed. The energy saving has been demonstrated to increase up to $70\%$ for a network containing up to $40$ nodes.}


\cite{joint_02} studies utility maximization problem subject to wireless network capacity and delay requirement of control system. The utility function is defined as the ratio of root-mean-square of the discrete-time system to that of the continuous-time counterpart. This utility function has been demonstrated to be a strictly concave function of the sampling period and inversely proportional to tracking error induced by discretization, based on the assumption that the plants follow the reference trajectories provided by the controllers. The wireless network capacity is derived by adopting slotted time transmission over a conflict graph, where each vertex represents a wireless link and there is an edge between two vertices if their corresponding links interfere with each other. The sampling period is used as the multihop end-to-end delay bound. The solution methodology is based on embedded-loop approach. In the inner loop, a relaxed problem with fixed delay bound, independent of sampling period, is solved via dual decomposition. The outer loop then determines optimal delay bounds based on the sampling period as an output of the inner loop.

\Blue{\cite{Alur11} proposes a mathematical framework for modeling and analyzing multihop NCSs. The authors present the formal syntax and semantics for the dynamics of the composed system, providing an explicit translation of multihop control networks to switched systems. The proposed method jointly considers control system, network topology, routing, resource scheduling, and communication error. The formal models are applied to analyze the robustness of NCSs, where data packet is exchanged through a multihop communication network subject to disruptions. The authors consider two communication models, namely, permanent error model and transient error model, dependent on the length of the communication disruptions. The authors address the robustness of the multihop NCS in the non-deterministic case by worst case analysis of scheduling, routing, and packet losses, and in the stochastic case by the stability analysis of node fault probability and packet loss probability.}

The joint optimization of the sampling period of sensors, packet forwarding policy and control law for computing actuator command is addressed in~\cite{burak_opt} for a multihop WirelessHART network. The objective of the optimization problem is to minimize the closed-loop control cost subject to the energy and delay constraints of the nodes. The linear quadratic cost function is used as the control cost similar to the one in~\cite{joint_03}. The solution methodology is based on the separation of joint design problem for the fixed sampling rate: transmission scheduling for maximizing the deadline-constrained reliability subject to a total energy budget and optimal control under packet loss. The optimal solution for transmission scheduling is based on dynamic programming, which allows nodes to find their optimal forwarding policy based on the statistics of their outgoing links in a distributed fashion. The bounds on the continuous-time control loss function are derived for optimal time-varying Kalman filter estimator and static linear feedback control law. The joint optimal solution is then found by a one-dimensional search over the sampling period.


\Blue{Some recent researches of WNCS investigate fault detection and fault tolerant issues~\cite{Ding13, Wang14}. \cite{Ding13} develops a design framework of fault-tolerant NCSs for industrial automation applications. The framework relies on an integrated design and parametrization of the TDMA MAC protocols, the controller, and the fault diagnosis algorithms in a multilayer system. The main objective is to determine the data transmission of wireless networks and reduce the traffic load while meeting the requirements of the control and the fault detection and identification performance. By considering the distributed control groups, the hierarchical WNCS configuration is considered. While the lower layer tightly integrates with sensors, actuators and microprocessors of (local) feedback control loops and its TDMA resource, the higher layer implements a fault-tolerant control in the context of resource management. The TDMA MAC protocol is modeled as a scheduler, whose design and parameterization are achieved with the development of the control and the fault detection and identification algorithms at the different functional layers. }



\Blue{In a similar way,~\cite{Wang14} investigates the fault estimation problem based on the deterministic model of the TDMA mechanism. The discrete periodic model of control systems is integrated with periodic information scheduling model without packet collisions. By adopting the linearity of state equations, the fault estimator is proposed for the periodic system model with arbitrary sensor inputs. The fault estimation is obtained after solving a deterministic quadratic minimization problem of control systems by means of recursive calculation. However, the scheduler of the wireless network does not consider any realistic message delays and losses.}


\textit{Routing and Traffic Generation Control:} In~\cite{Park16_cloc}, the cross-layer optimized control (CLOC) protocol is proposed for minimizing the worst-case performance loss of multiple control systems. CLOC is designed for a general wireless sensor and actuator network where both sensor--controller and controller--actuator connections are over a multihop mesh network. The design approach relies on a constrained max-min optimization problem, where the objective is to maximize the minimum resource redundancy of the network and the constraints are the stability of the closed-loop control systems and the schedulability of the communication resources. The stability condition of the control system has been formulated in the form of stochastic MATI constraint~\cite{control_bounded2}. The optimal operation point of the protocol is automatically set in terms of the sampling period, slot scheduling, and routing, and is achieved by solving a linear programming problem, which adapts to system requirements and link conditions. The performance analysis shows that CLOC ensures control stability and fulfills communication constraints while maximizing the worst-case system performance.

\Blue{\cite{Li16} presents a case study on a wireless process control system that integrates the control design and the wireless routing of the WirelessHART standard. The network supports two routing strategies, namely, single-path source routing and multi-path graph routing. Remind that the graph routing of the WirelessHART standard reduces packet loss through path diversity at the cost of additional overhead and energy consumption. To mitigate the effect of packet loss in the WNCS, the control design integrates an observer based on an extended Kalman filter with a MPC and an actuator buffer of recent control inputs. The experimental results show that sensing and actuation can have different levels of robustness to packet loss under this design approach. Specifically, while the plant state observer is highly effective in mitigating the effects of packet loss from the sensors to the controller, the control performance is more sensitive to packet loss from the controller to the actuators despite the buffered control inputs. Based on this observation, the paper proposes an asymmetric routing configuration for sensing and actuation (source routing for sensing and graph routing for actuation) to improve control performance.} 



\cite{Saifullah14_rate} addresses the sampling period optimization with the goal of minimizing overall control cost while ensuring end-to-end delay constraints for a multihop WirelessHART network. The linear quadratic cost function is used as the control performance measure, which is a function of the sampling period. The optimization problem relies on the multihop problem formulation of the end-to-end delay bound in~\cite{separate_08}. Due to the difficulty of the resulting optimization problem, the solution methodologies based on a subgradient method, simulated annealing-based penalty method, greedy heuristic method and approximated convex optimization method are proposed. The tradeoff between execution time and achieved control cost is analyzed for these methods.





\subsubsection{Event-Triggered Sampling}
\Blue{The communication system design for event-triggered sampling has mostly focused on the MAC layer. In particular, most researches focus on contention-based random access since it is suitable for these control systems due to the unpredictability of the message generation time.}

\textit{\Blue{Contention-based Access}:} \Blue{The tradeoff between the level threshold crossings in the control system and the packet losses in the communication system have been analyzed in~\cite{Maben09, demirel13, Vilgelm16, Gatsis16, Ramesh16, Mamduhi14, Molin14}. \cite{Maben09} studies the event-triggered control under lossy communication. The information is generated and sent at the level crossings of the plant output. The packet losses are assumed to have a  Bernoulli distribution independent over each link. The dependence between the stochastic control criterion on the level crossings and the message loss probability is derived for a class of integrator plants. This allows the generation of a design guideline on the assignment of the levels for the optimal usage of communication resources.}

\Blue{\cite{demirel13} provides an extension to~\cite{Maben09} by considering a multi-dimensional Markov chain model of the attempted and successful transmissions over lossy channel. In particular, a threshold-based event-triggering algorithm is used to transmit the control command from the controller to the actuator. By combining the communication model of the retransmissions with an analytical model of the closed-loop performance, a theoretical framework is proposed to analyze the tradeoff between the communication cost and the control performance and it is used to adapt an event threshold. However, the proposed Markov chain only considers the packet loss as a  Bernoulli process and it does not capture the contention between multiple nodes. On the other hand, schedule-based access, in which the nodes are assigned fixed time slots independent of their message generation times, is considered as an alternative to random access for event-triggered control~\cite{Jose14}. However, this introduces extra delay between the triggering of an event and a transmission in its assigned slot. }  

\Blue{\cite{Vilgelm16} analyzes the event-based NCS consisting of multiple linear time-invariant control systems over a multichannel slotted ALOHA protocol. The multichannel slotted ALOHA system is considered as the random access model of the Long Term Evolution~\cite{Tyagi15}. The authors separate the resource allocation problem of the multichannel slotted ALOHA system into two problems, namely, the transmission attempt problem and the channel selection problem. Given a time slot, each control loop decides locally whether to attempt a transmission based on some error thresholds. A local threshold-based algorithm is used to adapt the error thresholds based on the knowledge of the network resource. When the control loop decides to transmit, then it selects one of the available channels in uniform random fashion.}

\Blue{Given plant and controller dynamics,~\cite{Gatsis16} proposes control-aware random access policies to address the coupling between control loops over the shared wireless channel. In particular, the authors derive a sufficient mathematical condition for the random access policy of each sensor so that it does not violate the stability criterion of other control loops. The authors only assume the packet loss due to the interference between simultaneous transmissions of the network. They propose a mathematical condition decoupling the control loops. Based on this condition, a control-aware random access policy is proposed by adapting to the physical plant states measured by the sensors online. However, it is still computationally challenging to verify the condition.}

\Blue{Some event-triggered sampling appproaches~\cite{Ramesh16, Mamduhi14, Molin14} use the CSMA protocol to share the network resource. \cite{Ramesh16} analyzes the performance of the event-based NCSs with the CSMA protocol to access the shared network. The authors present a Markov model that captures the joint interactions of the event-triggering policy and a contention resolution mechanism of CSMA. The proposed Markov model basically extends Bianchi's analysis of IEEE 802.11~\cite{Bianchi00} by decoupling interactions between multiple event-based systems of the network.}

\Blue{\cite{Mamduhi14} investigates the event-triggered data scheduling of multiple loop control systems communicating over a shared lossy network. The proposed error-dependent scheduling scheme combines deterministic and probabilistic approaches. This scheduling policy deterministically blocks transmission requests with lower errors not exceeding predefined thresholds. Subsequently, the medium access is granted to the remaining transmission requests in a probabilistic manner. The message error is modeled as a homogeneous Markov chain. The analytical uniform performance bounds for the error variance is derived under the proposed scheduling policy. Numerical results show a performance improvement in terms of error level with respect to the one with periodic and random scheduling policies.}

\Blue{\cite{Molin14} proposes a distributed adaptation algorithm for an event-triggered control system, where each system adjusts its communication parameter and control gain to meet the global control cost. Each discrete-time stochastic linear system is coupled by the CSMA model that allows to close only a limited number of feedback loops in every time instant. The backoff intervals of CSMA are assumed to be exponentially distributed with homogeneous backoff exponents. Furthermore, the data packets are discarded after the limited number of retransmission trials. The individual cost function is defined as the linear quadratic cost function. The design objective is to find the optimal control laws and optimal event-triggering threshold that minimize the control cost. The design problem is formulated as an average cost Markov Decision Process (MDP) problem with unknown global system parameters that are to be estimated during execution. Techniques from distributed optimization and adaptive MDPs are used to develop distributed self-regulating event-triggers that adapt their request rate to accommodate a global resource constraint. In particular, the dual price mechanism forces each system to adjust their event-triggering thresholds according to the total transmission rate.}

\textit{Self-triggered Control and Mixed Approach:}  Self-triggered sampling allows to save energy consumption and reduce the contention delay by predicting the level crossings in the future, so, explicitly scheduling the corresponding transmissions~\cite{Jose14, self-control-carlo, Cao13, Henriksson15}. The sensor nodes are set to sleep mode until the predicted level crossing. \cite{self-control-carlo} proposes a new approach to ensure the stability of the controlled processes over a shared IEEE 802.15.4 network by self-triggered control. The self-triggered sampler selects the next sampling time as a function of current and previous measurements, measurement time delay, and estimated disturbance. The superframe duration and transmission scheduling in the contention free period of IEEE 802.15.4 are adapted to minimize the  energy consumption while meeting the deadlines. The joint selection of the sampling time of processes, protocol parameters and scheduling allows to address the tradeoff between closed-loop system performance and network energy consumption. However, the drawback of this sampling methodology is the lack of its robustness to uncertainties and disturbances due to the predetermined control and communication models. The explicit scheduling for self-triggered sampling is, therefore, recently extended to include additional time slots in the communication schedule not assigned apriori to any nodes~\cite{Jose14}. In the case of the presence of disturbance, these extra slots are used in an event-triggered fashion. The contention-based random access is used in these slots due to the unpredictability of the transmissions. 

\Blue{In~\cite{Cao13}, a joint optimization framework is presented, where the objective is a function of process state, cost of the actuations, and energy consumption to transmit control commands, subject to communication constraints, limited capabilities of the actuators, and control requirements. While the self-triggered control is adopted, with the controller dynamically determining the next task execution time of the actuator, including command broadcasting and changing of action, the sensors are assumed to perform sampling periodically. A simulated annealing based algorithm is used for online optimization, which optimizes the sampling intervals. In addition, the authors propose a mechanism for estimating and predicting the system states, which may not be known exactly due to packet losses and measurement noise. }

\Blue{\cite{Henriksson15} proposes a joint design approach of control and adaptive sampling for multiple control loops. The proposed method computes the optimal control signal to be applied as well as the optimal time to wait before taking the next sample. The basic idea is to combine the concept of the self-triggered sampling with MPC, where the cost function penalizes the plant state and control effort as well as the time interval until the next sample is taken. The latter is considered to generate an adaptive sampling scheme for the overall system such that the sampling time increases as the system state error goes to zero. In the multiple loop case, the authors also present a transmission scheduling algorithm to avoid the conflicts.}

\Blue{\cite{Peng16} proposes a mixed self-triggered sampling and event-triggered sampling scheme to ensure the control stability of NCSs, while improving the energy efficiency of the IEEE 802.15.4 wireless networks. The basic idea of the mixed approach is to combine the self-triggered sampling and the event-triggered sampling schemes. The self-triggered sampling scheme first predicts the next activation time of the event-triggered sampler when the controller receives the sensing information. The event-triggered sampler then begins to monitor the predefined triggering condition and computes the next sampling instance. Compared to the typical event-triggered sampling, the sensor does not continuously check the event-triggered condition, since the self-triggered sampling component of the proposed mixed scheme estimates the next sampling a priori. Furthermore, compared with the alone utilization of self-triggered sampling, the conservativeness is reduced, since the event-triggered sampling component extends the sampling interval. By coupling the self-triggered and event-triggered sampling in a unified framework, the proposed scheme extends the inactive period of the wireless network and reduces the conservativeness induced by the self-triggered sampling to guarantee the high energy-efficiency while preserving the desired control performance.}


\section{Experimental Testbeds}\label{sec:test}
\Blue{In contrast to previous surveys of WSN testbeds~\cite{Steyn11, Tonneau14, Horneber14}, we introduce some of our representative WNCS testbeds. Existing WNCS research often relies on small-scale experiments. However, they usually suffers from limited size, and cannot capture delays and losses of realistic large wireless networks. Several simulation tools~\cite{Cervin03,Eyisi12,Aminian13} are developed to investigate the NCS research. Unfortunately, simulation tools for control systems often lack realistic models of wireless networks that exhibit complex and stochastic behavior in real-world environments. In this section, we describe three WNCS testbeds, namely, cyber-physical simulator and WSN testbed, building automation testbed, and industrial process testbed.}



\subsection{Cyber-Physical Simulator and WSN Testbed}
\Blue{Wireless cyber-physical simulator (WCPS)~\cite{Li15} is designed to provide a realistic simulation of WNCS. WCPS employs a federated architecture that integrates Simulink for simulating the physical system dynamics and controllers, and TOSSIM~\cite{Levis03} for simulating wireless networks. Simulink is commonly used by control engineers to design and study control systems, while TOSSIM has been widely used in the sensor network community to simulate WSNs based on realistic wireless link models~\cite{Lee07}. WCPS provides an open-source middleware to orchestrate simulations in Simulink and in TOSSIM. Following the software architecture in WCPS, the sensor data generated by Simulink is fed into the WSN simulated using TOSSIM. TOSSIM then returns the packet delays and losses according to the behavior of the network, which are then fed to the controller of Simulink. Controller commands are then fed again into TOSSIM, which delays or drops the packets and sends the outputs to the actuators. Furthermore, it is also possible to use the experimental wireless traces of a WSN testbed as inputs to the TOSSIM simulator.}

\begin{figure}[]\centering
  \includegraphics[width = 0.95\columnwidth]{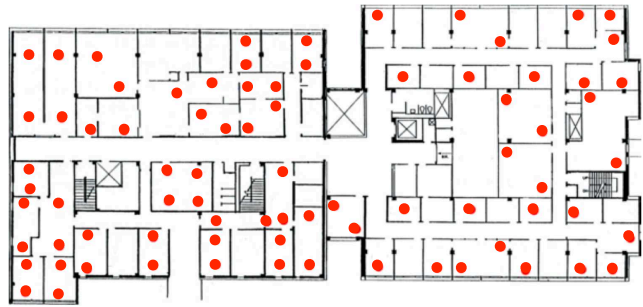}
  \caption{\Blue{WSN testbed in Bryan Hall and Jolley Hall of Washington University in St. Louis.}}
  \label{fig:wsan_test}
\end{figure}

\Blue{The Cyber-Physical Laboratory of Washington University in St. Louis has developed an experimental WSN testbed to study and evaluate WSN protocols~\cite{Sha15}. The system comprises a network manager on a server and a network protocol stack implementation on TinyOS and TelosB nodes~\cite{telos}. Each node is equipped with a TI MSP430 microcontroller and a TI CC2420 radio compatible with the IEEE 802.15.4 standard. Fig.~\ref{fig:wsan_test} shows the deployment of the nodes in the campus building. The testbed consists of 79 nodes placed throughout several office areas. The testbed architecture is hierarchical in nature, consisting of three different levels of deployment: sensor nodes, microservers, and a desktop class host/server machine. At the lowest tier, sensor nodes are placed throughout the physical environment in order to take sensor readings and/or perform actuation. They are connected to microservers at the second tier through a USB infrastructure consisting of USB 2.0 compliant hubs. Messages can be exchanged between sensor nodes and microservers over this interface in both directions. In the testbed, two nodes are connected to each microserver, typically with one microserver per room. The final tier includes a dedicated server that connects to all of the microservers over an Ethernet backbone. The server machine is used to host, among other things, a database containing information about the different sensor nodes and the microservers they are connected to.}



\subsection{Building Automation Testbed}
\begin{figure}[]\centering
  \includegraphics[width = 1\columnwidth]{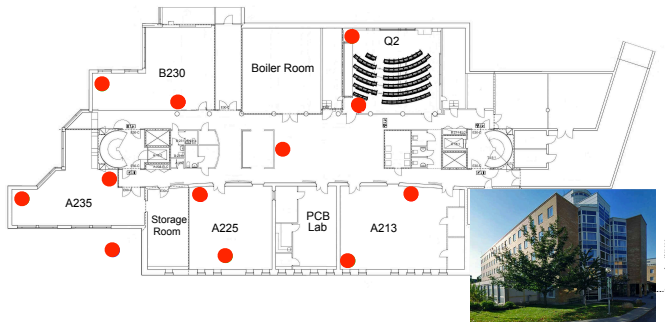}
  \caption{\Blue{HVAC testbed at the second floor of the Q-building at KTH. Each of the five rooms considered contain sensors and actuators used for HVAC control. Additional sensors are located in the corridor and outside of the building.}}
  \label{fig:hvac_test}
\end{figure}

\Blue{Heating, Ventilation and Air Conditioning (HVAC) systems guarantee indoor air quality and thermal comfort levels in buildings, at the price of high energy consumption~\cite{Aswani12}. To reduce the energy required by HVAC systems, researchers have been trying to efficiently use thermal storage capacities of buildings by proposing advanced estimation and control schemes by using wireless sensor nodes. An example HVAC testbed is currently comprised of the second floor of the electrical engineering building of the KTH campus and is depicted in Fig.~\ref{fig:hvac_test}. This floor houses four laboratories, an office room, a lecture hall, one storage room and a boiler room. Each room of the testbed is considered to be a thermal zone and has a set of wireless sensors and actuators that can be individually controlled. The WSN testbed is implemented on TinyOS and TelosB nodes~\cite{telos}. The testbed consists of 12 wireless sensors measuring indoor and outdoor temperature, humidity, CO2 concentrations, light intensity, occupancy levels, and events like door/windows openings/closings in several rooms. Note that the nodes are equipped with on-board humidity, temperature, and light sensors, and external sensors such as CO2 sensors by using an analog-to-digital converter channel on the 16-pin Telosb expansion area. Furthermore, laboratory A225 includes a people counter to measure the occupancy of the laboratory. The collection tree protocol is used to collect the sensor measurements through the multihop networks~\cite{Gnawali09}. The actuators are the flow valve of the heating radiator, the flow valve for the air conditioning system, the air vent for fresh air flow at constant temperature, and the air vent for air exhaust to the corridor.}

\begin{figure}[]\centering
  \includegraphics[width = 0.8\columnwidth]{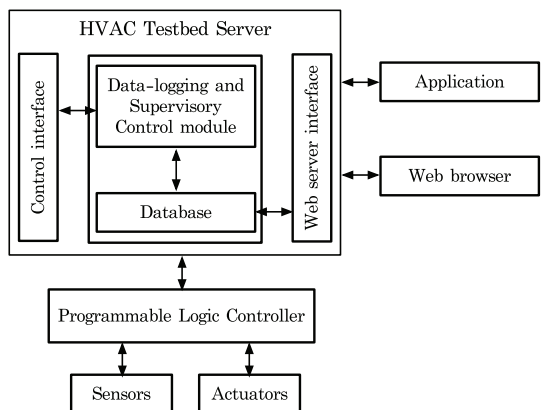}
  \caption{\Blue{HVAC system architecture. Users are able to design experiments through a LabVIEW application and remotely connect to the HVAC testbed. Additionally, through a web browser any user can download experimental data from the testbed database.}}
  \label{fig:hvac_ar}
\end{figure}

\Blue{An overview of the testbed architecture is shown in Fig.~\ref{fig:hvac_ar}. The HVAC testbed is developed in LabVIEW and is comprised of two separate components; the experimental application and a database/web server system~\cite{Pattarello13}. The database is responsible for logging the data from all HVAC components in real-time. On the other hand, the experimental application is developed by each user and interacts with the data-logging and supervisory control module in the testbed server, which connects to the programmable logic controller. This component allows for real-time sensing, computation, and actuation. Even though the application is developed in LabVIEW, MATLAB code is integrated in the application through a MathScript zone.}

%

\begin{figure}[]\centering
  \includegraphics[width = 1\columnwidth]{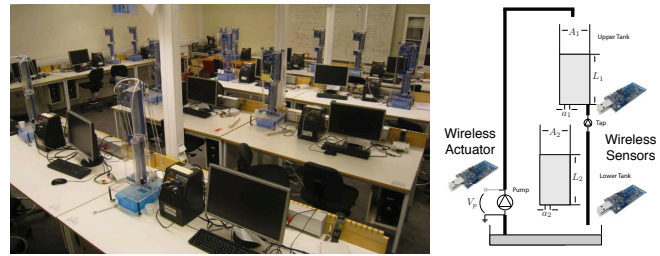}
  \caption{\Blue{Coupled tank system setup and its diagram.}}
  \label{fig:tank}
\end{figure}

\subsection{Industrial Process Testbed}
\Blue{The control of liquid levels in tanks and flows between tanks are basic problems in process industry~\cite{tank_book}. Liquids need to be processed by chemicals or mixed treatment in tanks, while the levels of the tanks must be controlled and the flows between tanks must be regulated. Fig.~\ref{fig:tank} depicts the experimental apparatus and a diagram of the physical system used in~\cite{Jose14}. The coupled tank system consists of a pump, a water basin and two tanks of uniform cross sections~\cite{Johansson00}. The system is simple, yet representative testbed of dynamics of water tanks used in practice. The water in the lower tank flows to the water basin. A pump is responsible for pumping water from the basin to the upper tank, which then flows to the lower tank. The holes in each of the tanks have the same diameter. The controller regulates the level of water in the upper or lower tank. The sensing of the water levels is performed by pressure sensors placed under each tank. The process control testbed is built on multiple control systems of Quanser coupled tanks~\cite{tank} with a wireless network consisting of TelosB nodes. The control loops are regulating two coupled tank processes, where the tanks are collocated with the sensors and actuators and communicate wirelessly with a controller node. A wireless node interfaces the sensors with an analog-to-digital converter, in order to sample the sensors for both tanks. The actuation is implemented through the digital-to-analog converter of the wireless actuator node, connected to an amplification circuit that will convert the output voltage of the pump motor. }



\vspace{10px}

\section{Open Challenges and \\ Future Research Directions} \label{sec:future}

%

Although a large number of results on WSN and NCSs are reported in the literature, there are still a number of challenging problems to be solved out, some of them are presented as follows.

\subsection{Tradeoff of Joint Design }
The joint design of communication and control layers is essential to guarantee the robustness, fault-tolerance, and resilience of the overall WNCS. Several different approaches of WNCS design are categorized dependent on the degree of the interaction. Increasing the interaction may improve the control performance but at the risk of high complexity of the design problem and thus eventually leading to the fundamental scalability and tractability issues. Hence, it is critical to quantify the benefit of the control performance and cost of the complexity depending on the design approaches. 

The benefit of the adaptation of the design parameters significantly depends on the dynamics of control systems. Most researches of control and communication focus on the design of the controller or the network protocol with certain optimization problems for the fixed sampling period. Some NCS researches propose possible alternatives to set the sampling periods based on the stability analysis~\cite{Branicky00, Zhang01, Zhang09}. However, they do not consider the fundamental tradeoff between QoS and sampling period of wireless networks. While the adaptive sampling period might provide control performance improvement, it results in the complex stability problem of the control systems and requires the real-time adaptation of wireless networks. Real-time adaptation of the sampling period might be needed for the fast dynamical system. On the other hand, it may just increase the complexity and implementation overhead for slow control systems. Hence, it is critical to quantify the benefit and cost of the joint design approach for control and communication systems.

\subsection{Control System Requirement}
Various technical approaches such as hybrid system, Markov jump linear system, and time-delay system are used to analyze the stability of NCSs for different network assumptions. The wireless network designers must carefully consider the detailed assumptions of NCS before using their results in wireless network design. Similarly, control system designers need to consider wireless network imperfections encompassing both message dropout and message delay in their framework. While some assumptions of control system design affect the protocol operation, other assumptions may be infeasible to meet for overall network. For instance, the protocol operation should consider the hard/soft sampling period to check whether it is allowed to retransmit the outdated messages over the sampling period. On the other hand, if the NCS design requires a strict bound on the maximum allowable number of consecutive packet losses, this cannot be achieved by the wireless system, in which the packet error probability is non-zero at all times.

Numerical methods are mostly used to derive feasible sets of wireless network requirements in terms of message loss probability and delay to achieve a certain control system performance. Even though all these feasible requirements meet the control cost, it may give significantly different network costs such as energy consumption and robustness and thus eventually affect the overall control systems. There are two ways to solve these problems. The first one is to provide efficient tools quantifying feasible sets and corresponding network costs. Previous researches of WNCS still lack of the comparison of different network requirements and their effect on the network design and cost. The second one is to provide efficient abstractions of both control and communication systems enabling the usage of non-numerical methods. For instance, the usage of stochastic MATI and MAD constraints for the control system in~\cite{Sadi14, sinem_icc} enables the generation of efficient solution methodologies for the joint optimization of these systems.

\subsection{Communication System Abstraction}
Efficient abstractions of communication systems need to be included to achieve the benefit of joint design while reducing complexity for WNCS. Both interactive and joint design approaches mostly focus on the usage of constant transmit power and rate at the physical layer to simplify the problem. However, variable transmit power and rate have already been supported by network devices. The integration of the variability of time slots with variable transmit power and rate has been demonstrated to improve the communication energy consumption significantly~\cite{Sadi13, sinem_twc}. This work should be extended to integrate power and rate variability into the WNCS design approaches.

Bernouilli distribution has been commonly used as a packet loss model to analyze the control stability for simplicity. However, most wireless links are highly correlated over time and space in practice~\cite{Beta08,Kappa10}. The time dependence of packet loss distribution can significantly affect the control system performance due to the effect of consecutive packet losses on the control system performance. The packet loss dependencies should be efficiently integrated into the interactive and joint design approaches.

\subsection{Network Lifetime} 
Safety-critical control systems must continuously operate the process without any interruptions such as oil refining, chemicals, power plants, and avionics. The continuous operation requires infrequent maintenance shut-downs such as semi-annual or annual since its effects of the downtime losses may range from production inefficiency and equipment destruction to irreparable financial and environmental damages. On the other hand, energy constraints are widely regarded as a fundamental limitation of wireless devices. The limited lifetime due to the battery constraint is particularly challenging for WNCS, because the sensors/actuators are attached to the main physical process or equipment. In fact, the battery replacement may require the maintenance shut-downs since it may be not possible to replace while the control process is operating. 

Recently, two major technologies of energy harvesting and wireless power transfer have emerged as a promising technology to address lifetime bottlenecks of wireless networks. Some of these solutions are also commercially available and deployed such as ABB WISA~\cite{wisa} based on the wireless power transfer for the industrial automation and EnOcean~\cite{enocean15} based on the energy harvesting for the building automation. WNCS using these energy efficient technologies encounters new challenges at all layers of the network design as well as the overall joint design approach. In particular, the joint design approach must balance the control cost and the network lifetime while considering the additional constraint on the arrival of energy harvesting. The timing and amount of energy harvesting may be random for the generation of energy from natural sources such as solar, vibration, or controlled for the RF, inductive and magnetic resonant coupling.

\subsection{Ultra-Reliable Ultra-Low Latency Communication}
\Blue{Recently, machine-type communication with ultra-reliable and ultra-low latency requirements has attracted much interest in the research community due to many control related applications in industrial automation, autonomous driving, healthcare, and virtual and augmented reality~\cite{Johansson15,Yilmaz15,Luvisotto16}. In particular, the Tactile Internet requires the extremely low latency in combination with high availability, reliability and security of the network to deliver the real-time control and physical sensing information remotely~\cite{Fettweis14}.} 

Diversity techniques, which have been previously proposed to maximize total data rate of the users, are now being adapted to achieve reliability corresponding to packet error probability on the order of $10^{-9}$ within latency down to a millisecond or less. The ultra-low latency requirement may prohibit the sole usage of time diversity in the form of automatic-repeat-request (ARQ), where the transmitter resends the packet in the case of packet losses, or hybrid ARQ, where the transmitter sends incremental redundancy rather than the whole packet assuming the processing of all the information available at the receiver. Therefore,~\cite{ultra_01, ultra_02, ultra_03, ultra_04} have investigated the usage of space diversity in the form of multiple antennas at the transmitter and receiver, and transmission from multiple base stations to the user over one-hop cellular networks. These schemes, however, mostly focus on the reliability of a single user~\cite{ultra_01, ultra_04}, multiple users in a multi-cell interference scenario~\cite{ultra_02}, or multiple users to meet a single deadline for all nodes~\cite{ultra_03}. \cite{ultra_05} extended these works to consider the separate packet generation times and individual packet transmission deadlines of multiple users in the high reliability communication.

The previous work on WNCS only investigated the time and path diversity to achieve very high reliability and very low latency communication requirements of corresponding applications, as explained in detailed above. The time diversity mechanisms either adopt efficient retransmission mechanisms to minimize the number of bits in the retransmissions at the link layer or determine the best timing and quantity of time slots given the link quality statistics. On the other hand, path diversity is based on the identification of multiple disjoint paths from source to destination to guarantee the routing reliability against node and link failures. The extension of these techniques to include other diversity mechanisms, such as space and frequency in the context of ultra-reliable ultra low latency communication, requires reformulation of the joint design balancing control cost and network lifetime and addressing new challenges at all layers of the network design. 


\subsection{Low-Power Wide-Area Networks}
One of the major issues for large scale Smart Grid~\cite{Yu16}, Smart Transportation~\cite{Zhang11}, and Industry 4.0~\cite{Ind40} is to allow long-range communications of sensors and actuators using very low-power levels. Recently, several LPWAN protocols such as  LoRa~\cite{LoRa}, NB-IoT~\cite{NB_IoT}, Sigfox~\cite{Sigfox}, and LTE-M~\cite{Condoluci16} are proposed to provide the low data rate communications of battery operated devices. LTE-M and NB-IoT use a licensed spectrum supported by 3rd Generation Partnership Project standardization. On the other hand, LoRa and Sigfox rely on an unlicensed spectrum.

The wireless channel behavior of LPWANs is significantly different from the behavior of the short-range wireless channel commonly used in WNCS standards, such as WirelessHART, Bluetooth, and Z-wave, due to different multi-path fading characteristics and spectrum usage. Thus, the design of the physical and link layers is completely different. Moreover, the protocol design needs to consider the effect of the interoperation of different protocols of LPWANs on the overall message delay. Hence, the control system engineers must validate the feasibility of the traditional assumptions of wireless networks for WNCS based on LPWANs. Furthermore, the network architecture of LPWAN must carefully adapt its operation in order to support the real-time requirements and control message priority of large scale control systems.

\section{Conclusions} 
Wireless networked control systems are the fundamental technology of the safety-critical control systems in many areas, including automotive electronics, avionics, building automation, and industrial automation. This article provided a tutorial and reviewed recent advances of  wireless network design and optimization for wireless networked control systems. We discussed the critical interactive variables of communication and control systems, including sampling period, message delay, message dropout, and energy consumption. We then discussed the effect of wireless network parameters at all protocol layers on the probability distribution of these interactive variables. Moreover, we reviewed the analysis and design of control systems that consider the effect of various subsets of interactive variables on the control system performance. By considering the degree of interactions between control and communication systems, we discussed two design approaches: interactive design and joint design. We also describe some practical testbeds of WNCS. Finally, we highlighted major existing research issues and identified possible future research directions in the analysis of the tradeoff between the benefit of the control performance and cost of the complexity in the joint design, efficient abstractions of control and communication systems for their usage in the joint design, inclusion of energy harvesting and diversity techniques in the joint design and extension of the joint design to wide-area wireless networked control systems.

\bibliographystyle{IEEEtran}
\bibliography{ref}

\end{document}